\documentstyle[12pt,epsf]{article}

\newtheorem{lemma}{Lemma}[section]
\newtheorem{proposition}{Proposition}[section]
\newtheorem{theorem}{Theorem}[section]
\newtheorem{corollary}{Corollary}[section]
\def\proof{\par{\it Proof}. \ignorespaces}
\def\endproof{{\ \vbox{\hrule\hbox{%
   \vrule height1.3ex\hskip0.8ex\vrule}\hrule }}\par}
\newenvironment{Proof}{\proof}{\endproof}

\def\pat{{\partial}}
\def\la{{\lambda}}

\def\sig{{\sigma}}
\def\be{{\beta}}
\def\om{{\omega}}

\title{The Whitham Equations for \\ Optical Communications: \\
Mathematical Theory of NRZ}
\author{Yuji Kodama\thanks{e-mail: kodama@comm.eng.osaka-u.ac.jp}\\
{}\\
\it Department of Communications Engineering \\
\it Osaka University,  Suita,  Osaka 565, Japan}
\begin{document}
\maketitle
\bibliographystyle{plain}

\begin{abstract}
We present a model of optical communication system for high-bit-rate data
transmission in the nonreturn-to-zero (NRZ) format over transoceanic distance.
The system operates in a small group velocity
dispersion regime, and the model equation
is given by the Whitham equations describing the slow modulation of multi-phase
wavetrains of the (defocusing) nonlinear Schr\"odinger (NLS) equation.
The model equation is of hyperbolic type, and certain initial NRZ pulse
with phase modulation
develops a shock.  We then show how one can obtain a global solution
by choosing an appropriate Riemann surface on which the Whitham equation
is defined.
The present analysis may be interpreted as an alternative
to the method of inverse scattering transformation for the NLS solitons.
We also discuss wavelength-division-multiplexing (WDM) in the NRZ format
by using the Whitham equation for a coupled NLS equation, and show that
there exists a hydro-dynamic-type instability between channels.

\end{abstract}

\section{Introduction}
\renewcommand{\theequation}{1.\arabic{equation}}\setcounter{equation}{0}

Recently there has been a great deal of research
works on designing a long-distance and
high-speed optical communication system for the next generation of 100Gbit/s
over 9000km (transpacific distance).  In the fall of
1996, the undersea optical cable
called TPC-5 between Japan and US has been installed, and the system
operates at 10Gbit/s in a non-return-to-zero (NRZ) format.
Also in 1998, a total of 100Gbit/s (e.g. 5Gbit/s $\times$ 20Channels)
wavelength-division-multiplexed (WDM)-NRZ
system will be set up as TPC-6 optical cable.  In such a system,
noise accumulation and the combined action of fiber dispersion in the
group velocity of pulse and fiber nonlinearity called the Kerr effect
are the main system limiting factors.  The analysis of NRZ signal propagation
was hampered so far by necessity of lengthy numerical simulation and expensive
experiments.  Recently, we proposed in \cite{kodama:95a}
a hydrodynamic model to describe nonlinear effects in
the NRZ pulse propagation, which was derived as a
weak dispersion limit of the nonlinear Schr\"odinger (NLS) equation.
Based on this model, we also discussed methods to control the pulse broadening
by means of initial phase modulation (pre-chirping) \cite{kodama:95b} and
nonlinear
gain \cite{kodama:96a}.

In this paper, we give a mathematical formulation of the model in
the framework of the Whitham averaging method \cite{whitham:78}
which describes the slow modulation of the amplitude and phase of
 quasi-periodic solution of the NLS equation.
We start in Section 2 to provide a necessary information of the problem,
and derive a shallow water wave equation as a simplest approximation of the
model.
In Section 3, we give a mathemtical background of the Whitham equation as
describing dynamics of the genus $g$ Riemann surface determined
by the spectral curve of the Lax operator for the NLS equation.
Here the genus represents the number of periods in the quasi-periodic
solution, and the Whitham equations are given by first order
quasi-linear system of partial differential equations.
In particular, we identify the shallow
water wave equation as the genus zero Whitham equation.  We also note
that the spectral curves corresponding to the Whitham equation for the
NLS equation (NLS-Whitham equation) are hyper-elliptic.
In Section 4, we first show the hyperbolicity
of the NLS-Whitham equations which enables us to classify the initial
data for the global existence of the solution.  We then
decribe the effect of pre-chirping for reducing
the broadening of NRZ pulse, and as a result we find that the Riemann
surface becomes singular (formation of a {\it shock}),
 and its genus changes from zero to
either one or two depending on the strength of the chirp.

 Since the NRZ system operates at a small group velocity (i.e. second
order) dispersion
 regime, the third order dispersion becomes important, especially for
pulses with
shorter width (in a higher bit rate system).  We discuss this effect in
Section 5,
and show that the third order dispersion leads to asymmetric distortion of
the NRZ pulse and results a formation of shock-like structure.  In Section 6,
we extend the NLS-Whitham model to describe pulse propagation in a WDM-NRZ
system.
The extension is obtained by taking a weak dispersion limit of a coupled
NLS equation, in which the interaction between channels due to the Kerr
nonlinearity is expressed by the cross phase modulation.  We then find a
hydrodynamic-type instability between channels, which corresponds to
loosing the hyperbolicity
of the hydrodynamic model obtained in the dispersionless
limit of this coupled NLS equations.
 The WDM model is
further studied by using an integrable model, the Manakov equation
\cite{manakov:78} or the
vector NLS equation.   The Whitham equation of genus zero
in this model is the well-known Benney equation which describes shallow water
waves in a stratified fluids \cite{benney:80,zakharov:80},
and the number of channels in
WDM corresponds to that of the layers.  We then present how the nonzero genus
Whitham equations appear in the Manakov equation.  The main feature of this
model
is that the spectral curve is non-hyperelliptic, and the corresponding generic
Riemann surface
has a topology of a sphere with $3h$ handles with some positive integer $h$.

\section{Preliminary}
\renewcommand{\theequation}{2.\arabic{equation}}\setcounter{equation}{0}

For a long-distance and high-bit rate optical communication
system, both dispersion and nonlinearity of the optical fiber are important
and the NLS equation has been
used as a model equation expressing such effects \cite{hasegawa:94},
\begin{eqnarray}
i {{\pat q} \over {\pat Z}} + {\beta_2 \over 2}
{{\pat^2 q} \over {\pat T^2}} + \nu
|q|^2 q & = & 0,
\label{nls}
\end{eqnarray}
Here $ q $ is the complex envelope function of the electric field in
the fiber, $
Z $ the propagation distance, and $ T $ the retarded time.  The
coefficients $ \beta_2 $ and $ \nu (>0) $ represent the group
velocity dispersion (GVD) and the Kerr
(nonlinear) effect of the fiber.  Here we further normalize $q$ so that
we can set $\nu=1$.  It is well known
that if $ \be_2 > 0 $ (focusing or modulationally unstable case),
the NLS equation has so-called bright solitons
with sech-shape, and if $ \be_2 < 0 $ (deforcusing or modulationally stable
case),
so-called dark
ones with tanh-shape. These are the stationary solutions for Eq.(\ref{nls}),
and Fig.\ref{f2:1} shows examples of pulse sequences in $|q|^2$ as
a 16bits-coded signal (0010110010111100) using those pulses.
\begin{figure}
\epsfysize=8cm
\centerline{\epsffile{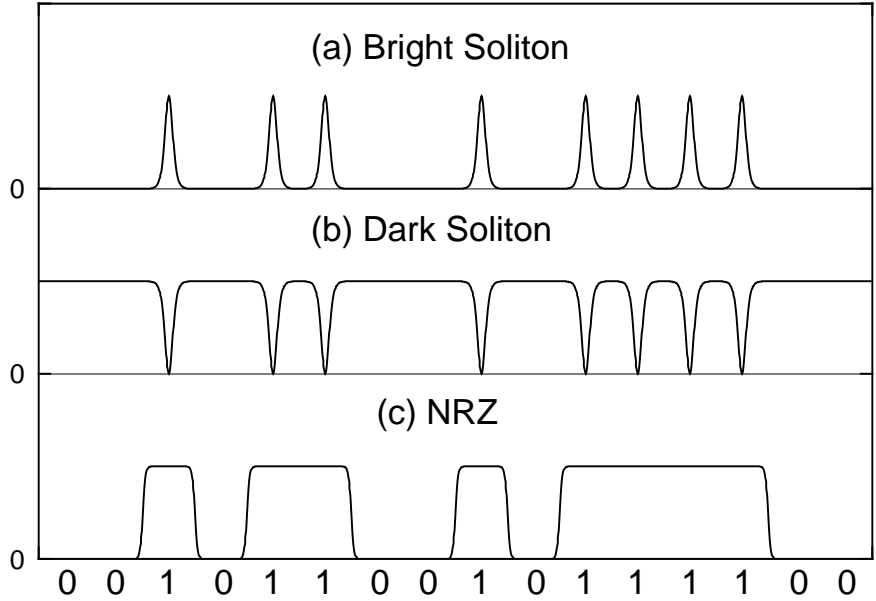}}
\caption{Various types of a 16bits coded signal (0010110010111100).}
\end{figure}
In Fig.\ref{f2:1}, we also show
the coded signal using the NRZ pulses.  Because of the modulational stability
of constant amplitude, for example, in the (011110) coding part,
one has to operate the NRZ system at the deforcusing regime, $\beta_2<0$.
The NRZ pulse is however non-stationary, and in order to reduce the pulse distortion
the system is designed in a small dispersion
and low power regime.  In an ideal case,
the best performance can be obtained in the zero dispersion and linear limits.
However, the real system always includes a noise, and the signal pulse power
should be kept more than the noise level.  Because of this limit in signal-noise
ratio (SNR), the nonlinearity becomes an essential effect for a long distance
transmission problem, and the zero dispersion causes a resonant interaction
process
between noise and signal, which is called the four wave mixing (FWM), and
leads to a distortion of the pulse.  The main purpose of this paper is to
develop
a mahematical theory to describe the pulse distortion in a weak dispersion
limit of the NLS equation.  Here we do not discuss, however, the distortion
due to the FWM with noise.

 Let us first recall that the NLS equation (\ref{nls})
 has an infinite number of conservation laws, for example,
\begin{eqnarray}
& & \frac{\partial}{\partial Z}|q|^2 = \frac{\partial}{\partial T}
\left(\beta^2|q|^2 \frac{\partial}{\partial T} \mbox{Arg}(q)\right),
\label{cons1} \\
& & \frac{\partial}{\partial Z}\left(|q|^2 \frac{\partial}{\partial T}
\mbox{Arg}(q)\right)= \frac{\partial}{\partial T} \left( -\frac{\beta^2}{4}
\frac{\partial^2 q}{\partial T^2}
+ \beta^2\left|\frac{\partial q}{\partial T}\right|^2 + \frac{1}{2}|q|^4
\right). \label{cons2}
\end{eqnarray}
Here we have set $ \beta_2=- \be^2 $.  The quantities $|q|^2$ and Arg($q$)$:=
(i/2)\ln (q^*/q)$ represent the local intensity and phase of the electric field.
 We then introduce $\rho:=|q|$ and $\sigma:= \beta$Arg($q$), and rewrite $
q $ as
\begin{equation}
\label{q}
 q(T,Z)=\sqrt {\rho (T,Z)} \exp\left(i \frac{1}{\beta} \sig (T,Z) \right).
\end{equation}
  Substituting Eq.(\ref{q}) to Eqs.(\ref{cons1}) and (\ref{cons2}), and
rescaling the variables $ Z \rightarrow Z / \be $,  we have
\begin{eqnarray}
& & \frac{\partial}{\partial Z}\rho = \frac{\partial}{\partial T}
\left(\rho u \right), \label{crhou1} \\
& & \frac{\partial}{\partial Z}\left(\rho u \right)= \frac{\partial}{\partial T}
 \left( \rho u^2 + \frac{1}{2}\rho^2 -\frac{\beta^2}{4} \rho\frac{\partial^2}{
 \partial T^2} \ln \rho \right). \label{crhou2}
\end{eqnarray}
where $u$ represents the chirp defined by $u:=\partial \sigma/\partial T$.
In a small dispersion limit $ \be \rightarrow 0 $, if we assume that both
$\rho$ and
$u$ are smooth, then Eqs.(\ref{crhou1}) and (\ref{crhou2})
for $\rho \ne 0$ can be approximated by a $2 \times 2$ quasi-linear
system,
\begin{eqnarray}
{\pat \over {\pat Z}} \left(\matrix {\rho \cr  u \cr}
\right) & = & \left(\matrix { u & \rho \cr  1 &  u
\cr} \right) {\pat \over {\pat T}} \left(\matrix {\rho \cr
u \cr} \right).  \label{g0}
\end{eqnarray}
A detail on the weak dispersion limit of the NLS equation
has been studied in \cite{jin:95}, where the existence theorem of the limit
is given based on the Lax-Levermore theory.
When $  \rho > 0 $ in (\ref{g0}), the eigenvalues of
the coefficient matrix are real, $  u + \sqrt {
\rho} $ and $  u - \sqrt { \rho} $, and  the system is totally
hyperbolic.  The system (\ref {g0}) is known as the shallow
water wave equation, and has been intensively discussed (for
example, see \cite{whitham:78}).  It is then interesting to note that the
distortion of the NRZ pulse may be understood as a deformation
of the water surface.

For a demonstration of the NRZ pulse propagation, we
 consider the initial value problem of Eq.(\ref{g0}).
As a simple example of the initial NRZ pulse, we take a
square pulse having constant phase (zero-chirp),
\begin{eqnarray}
 \rho (T,0)  =  \left \{\begin{array} {ll}
                                 \rho_0,  & \mbox{for~~ $ |T| < T_0 $} \\
                                 0, & \mbox{for~~ $ |T| > T_0 $}
                                \end{array} \right . \ , ~~~~~~
 u (T,0) = 0, \ \ \mbox{for~~ $\forall T$}.
 \label{ig0}
\end{eqnarray}
This is called "Dam-breaking problem", since $\rho$ and $-u$ represent the depth
and the velocity of water which rests on the {\it spatial region} $|T|<T_0$ at
the {\it time} $Z=0$.  We then expect to see a leakage of the water from the
edges, and in fact
we obtain:
\begin{proposition}
\label{prop1}
The system (\ref {g0}) with the initial data
(\ref {ig0}) has the following global solution
up to the distance $Z=T_0/\sqrt{\rho_0}$: For $T>0$,
\begin{eqnarray*}
\rho (T,Z)&=&u(T,Z)=0, \ \ ~~~~~~~~~~~ \mbox{for} \ \
T>T_0+2\sqrt{\rho_0}Z, \cr
 & & {} \cr
\rho(T,Z) &=& \mbox{min}\left[\rho_0, \frac{1}{9}\left(2\sqrt{\rho_0}
-\frac{T-T_0}{Z}\right)^2\right], \ \  \mbox{for } \ \
0<T<T_0+2\sqrt{\rho_0}Z, \cr
 & & {} \cr
u(T,Z)&=& - \mbox{max}\left[0,{2 \over 3} \left (  \sqrt {\rho_0}+
{T-T_0 \over Z} \right )\right], \ \ ~
 \mbox{for } \ \ 0<T<T_0+2\sqrt{\rho_0}Z,
\end{eqnarray*}
For $T<0$, we have $\rho(T,Z)=\rho(-T,Z)$ and $u(T,Z)=-u(-T,Z)$.
\end{proposition}
\begin{Proof}
 As in the standard analysis of the method of characteristics for the
quasi-linear hyperbolic system (\ref{g0}), we
first rewrite the system in the Riemann invariant form \cite{whitham:78}
\begin{eqnarray}
{\pat \over {\pat Z}} \left(\matrix {\la_1 \cr \la_2 \cr} \right ) & =
& {1 \over 4} \left (\matrix {3 \la_1 + \la_2 & 0 \cr 0 & \la_1 + 3
\la_2 \cr} \right ) {\pat \over {\pat T}} \left (\matrix {\la_1 \cr
\la_2 \cr} \right ), \label{rg0}
\end{eqnarray}
where the Riemann invariants $ \la_1 $ and $ \la_2 $ are given by
\begin{equation}
\la_1 =  u - 2 \sqrt {\rho}, ~~~~~  \la_2 =  u + 2
\sqrt { \rho}. \label{l0}
\end{equation}
Then the initial data for the Riemann invariants are given by
\begin{eqnarray}
& & \la_1(T,0) =  \left \{\begin{array} {ll}
             2 \sqrt {\rho_0},  & \mbox{for~~~ $ T < -T_0 $} \\
             -2\sqrt {\rho_0},  & \mbox{for~~~ $ T > -T_0 $}
             \end{array} \right .  ,\nonumber \\
 & & {} \label{initial0} \\
& & \la_2(T,0) =  \left \{\begin{array} {ll}
             2 \sqrt {\rho_0},  & \mbox{for~~~ $ T < T_0 $} \\
             -2\sqrt{\rho_0}, & \mbox{for~~~ $ T > T_0 $}
             \end{array} \right . . \nonumber
\end{eqnarray}
Here we extend the values of $\lambda_1$ and $\lambda_2$ for the region
$|T|>T_0$.  One should note from Eq.(\ref{crhou2})
that $u$ is defined only the region where $\rho\ne 0$.  Then Eq.(\ref{rg0})
becomes a single equation for $\lambda_2$ near $T=T_0$ where $\lambda_1=
-2\sqrt{\rho_0}$,
\begin{equation}
\frac{\partial \lambda_2}{\partial Z}=\frac{1}{4} (-2\sqrt{\rho_0}+
3\lambda_2)\frac{\partial \lambda_2}{\partial T},
\label{l2}
\end{equation}
from which we have the solution expressing a rarefaction (centered simple) wave
starting from $T=T_0$; for $T>0$,
\begin{equation}
\la_2(T,Z)=
\left \{\begin{array} {lll}
      \displaystyle{2 \sqrt {\rho_0}},
       & \mbox{for~~~ $ 0 <T < T_0-\sqrt{\rho_0}Z $}, \\
      \displaystyle{ \frac{2}{3} \sqrt{\rho_0}-\frac{4}{3}\frac{T-T_0}{Z}},
      & \mbox{for~~ $ T_0-\sqrt{\rho_0}Z< T < T_0+2\sqrt{\rho_0}Z $}, \\
      \displaystyle{ -2\sqrt{\rho_0}},
     & \mbox{for~~~ $ T > T_0+2\sqrt{\rho_0}Z $}.
             \end{array} \right .
\label{hod}
\end{equation}
This solution is valid up to the distance $Z=T_0/\sqrt{\rho_0}$ where
 the centered simple waves from $T=T_0$ and
$-T_0$ collide at $T=0$, and the solution should be modified.  To find
$\la_1(T,Z)$,
note that the point symmetry in the graphs ($\la_k, T$), i.e.
$\la_2(T)=-\la_1(-T)$.
Then from $\la_1$ and $\la_2$ we have the solutions
\begin{equation}
\label{rou}
\rho=\frac{1}{16}(\la_1-\la_2)^2, ~~~~~~ u=\frac{1}{2}(\la_1+\la_2).
\end{equation}
Noting the symmetry $\rho(-T,Z)=\rho(T,Z)$ and $u(-T,Z)=-u(T,Z)$, we complete
the proof.
\end{Proof}
\begin{figure}
\epsfysize=7.5cm
\centerline{\epsffile{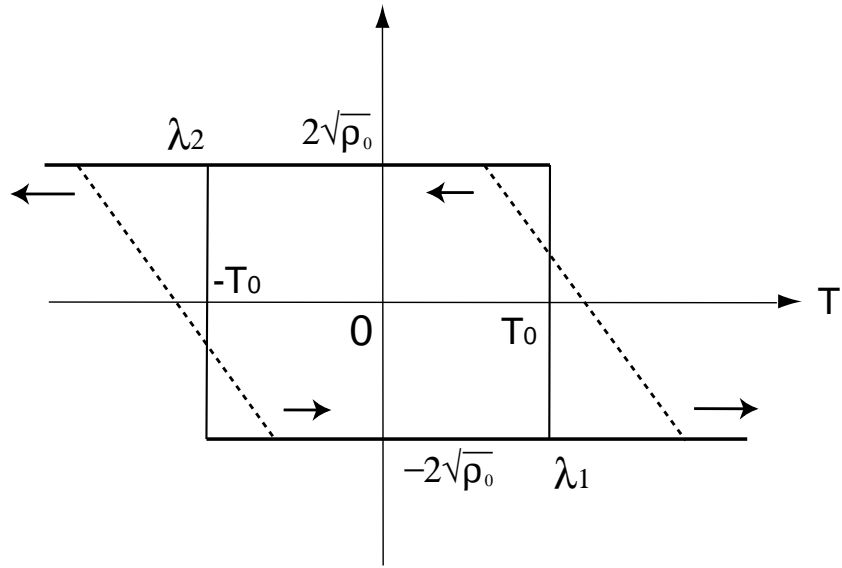}}
\caption{Evolution of the Riemann invariants.  Dashed curves are
the initial data: Solid curves are the solution $\la_k(T,Z)$ for $Z>0$.}
\end{figure}
\begin{figure}
\epsfysize=6cm
\centerline{\epsffile{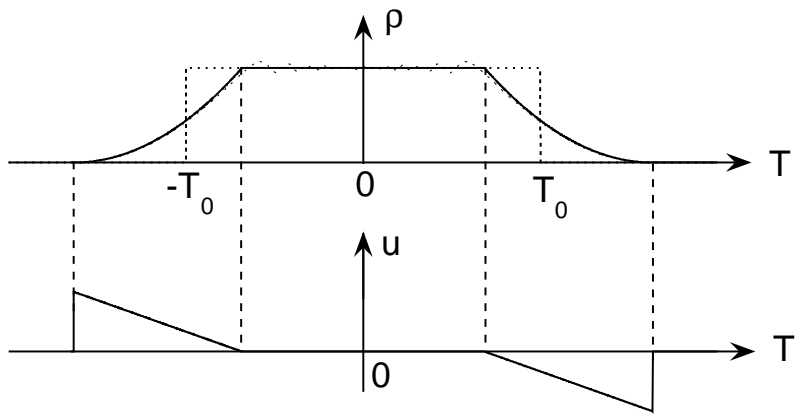}}
\caption{Deformation of NRZ pulse (Dam breaking problem). The thin dotted curve
on $\rho$ shows the numerical solution of the NLS equation.}
\end{figure}
\vspace{0.2cm}

The evolution of the Riemann invariants $\la_1$ and $\la_2$
is shown in Fig.\ref{f2:2}.
In Fig.\ref{f2:3}, we plot the pulse shape $\rho$ and the chirp
$u$ obtained in Proposition \ref{prop1}.   The thin dotted curve in
Fig.\ref{f2:3}
shows the numerical result of the NLS equation with $\rho_0=1, T_0=10$ and
$\beta_2\equiv -\beta^2=-0.1$ at $Z=10\beta\approx 3.16$.  Notice the good
agreement with the analytical solution except some small oscillations on the top
of $\rho$ which disappear in the limit $\beta_2\to 0$.
As we have predicted
from the hydrodynamic analogy, the deformation of the pulse induces the
generation of chirp (or water velocity) at the edges, and the global nature of
the solution is understood as an expansion of the water (the rarefaction wave).
The expansion is characterized by the characteristic velocity $s$ at the edge,
and for example at $T=T_0$ it is given by $s=-\sqrt{\rho_0}$ for $T<T_0$ and
$s=2\sqrt{\rho_0}$ for $T>T_0$.
In order to reduce the expansion, it is natural to put initial chirp
opposed to the chirp appearing in the edges.  This is equivalent to give an
initial velocity with a piston and to compress the water.
Because of the quasi-linearity of
the system (\ref{g0}), we then expect a shock formation in the solution, and
therefore Eq.(\ref{g0}) is no longer valid as an approximate
model of the NLS equation.
In fact, if we start with the initial data with the same $\rho$ in (\ref{ig0})
but nonzero chirp $u$, for example, with $u_0>0$
\begin{equation}
u(T,0) =  \left \{\begin{array} {ll}
             -u_0, ~~ & \mbox{for~~~ $ T < 0 $} \\
             u_0, ~~ & \mbox{for~~~ $ T>0 $}
             \end{array} \right . , \label{initialu}
\end{equation}
the numerical solution of the NLS equation results high oscillations
starting from the
discontinuous point of chirp, $T=0$, as seen in Fig.\ref{f2:4}.
\begin{figure}
\epsfysize=7cm
\centerline{\epsffile{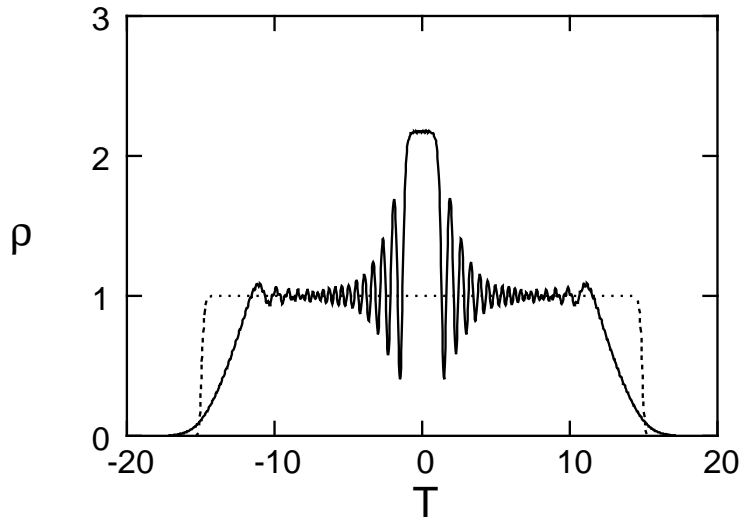}}
\caption{Optical shock due to initial chirp.}
\end{figure}
Here we have set $\rho_0=1, u_0=3\beta\approx 0.95$ with $\beta_2=-0.1$,
and the solution $\rho(T,Z)$ is at $Z=5\beta\approx 1.58$.
This phenomena has been also noted in \cite{bronski:94} as an optical
shock appearing in somewhat similar situation.
The main frequency of the oscillations
is of order $1/\beta$, and usual optical filter can remove
those oscillations as in the sense of {\it averaging}.
So what we want to describe here is the average behavior of the solution.
For this purpose, in the following section
we extend the model (\ref{g0}) and show that
the new model admits a global solution describing the average
motion of the NLS equation (\ref{nls}) for several step initial data
having different values of chirp.  This is the main purpose of the present
paper, and it
may be referred as a dispersive regularization problem of the shock
singularity of hyperbolic system in a weak dispersion limit of the NLS
equation.

\section{The NLS-Whitham equations}
\renewcommand{\theequation}{3.\arabic{equation}}\setcounter{equation}{0}

We here develop a regularization method based on the integrability of
the NLS equation.  The key idea is to identify the shock singularity as
a generation of a quasi-periodic solution of the NLS equation which appears as
the high oscillations as seen in Fig.4.  The quasi-periodic solution can be
characterized by the Riemann surface defined by the
spectral curve of the Lax operator in the inverse scattering
transform (IST).  Then the genus of the Riemann surface
 changes through the singularity.  Here the number of the genus
corresponds to that of independent phases in the quasi-periodic solution
\cite{forest:88,novikov:80}.  Then the main purpose here is to derive an
 equation to describe the dynamics of the Riemann surface corresponding to
 the modulation of the quasi-periodic solution of the NLS equation.  The
resulting equation is called the NLS-Whitham equation.

Let us first recall the IST scheme of the
NLS equation and reformulate the quasi-linear system (\ref {g0}) as
the simplest approximation of the NLS equation in a weak dispersion
limit, that is, the genus zero NLS-Whitham equation.  Then we discuss the \
general case with arbitrary genus.
Here we somewhat use a different formulation of the IST method, which is
related to Sato's formulation of the KP hierarchy \cite{sato:80}.  This is quite
useful not only for the present purpose but also for an integrable extension
of coupled NLS equation used as a model of WDM system discussed in Section 6.
The Lax operator $ L $ of the NLS equation in this formulation is given by a
pseudo-differential operator with $ D := - i \beta \pat / \pat T $,
\begin{eqnarray}
L & = & D + \sqrt {\rho} (D - u)^{-1} \sqrt {\rho} \nonumber \\
& = & D + \rho D^{-1} + \left (\rho u - {1 \over 2} D \rho \right ) D^{-2} +
\cdots  \  . \label{l}
\end{eqnarray}
This can be derived from the standard Lax operator in a $2\times 2$ matrix
form \cite{zakharov:71}, and the operation with $D^{\ell}$ is given by a
generalized Leibnitz rule with $DD^{-1}=D^{-1}D=1$,
\begin{equation}
\label{leipnitz}
D^{\ell}f\cdot=\sum_{k=0}^{\infty} \left(\matrix{\ell \cr k \cr}\right)(D^kf)
D^{\ell-k}\cdot \ \ , \ \ \mbox{for \ any} \ \ell\in {\bf Z}.
\end{equation}
Then the IST scheme (the Lax pair) for the NLS equation is given by
 the eigenvalue problem $ L\phi = \la \phi $ and the evolution equation,
\begin{equation}
-i {{\pat \phi} \over {\pat Z}} = B_2 \phi, ~~~~ B_2 = {1 \over 2}
(L^2)_{\ge 0} :=
{1 \over 2} D^2 + \rho. \label{kp}
\end{equation}
Here the symbol $(L^2)_{\ge 0}$ denotes the differential (i.e. polynomial
in $D$)
part of the pseudo-differential operator $L^2$.

 As a genelarization of this scheme, the integrable hierarchy of the
NLS equation can be formulated as
\begin{equation}
-i{{\pat \phi} \over {\pat Z_n}} = B_n \phi, ~~~~ B_n = {1 \over n}
(L^n)_{\ge 0},
 \label{hkp}
\end{equation}
where $Z_n$ is the parameter describing the evolution generated by $B_n$ for
$n=1,2,\cdots$.  In particular, $Z_1$ represents the translation in $T$,
that is,
$\phi(T,Z_1,Z_2,\cdots)=\phi(T+\beta Z_1,Z_2,\cdots)$ so that we identify
$\beta Z_1$ as $T$, and $Z_3$ represents the
parameter for the complex modified KdV equation which will play an
important rule in
Section 5 where we discuss a higher order correction to the NLS equation.
The compatibility conditions for the $\phi$-equations, $L\phi=\lambda\phi$
and (\ref{hkp}), then give the Lax equations
\begin{equation}
 -i {{\pat L} \over {\pat Z_n}} = [B_n, L]:=B_nL-LB_n, \label{lax}
\end{equation}
from which we recover Eqs.(\ref {crhou1}) and (\ref{crhou2}) for $n=2$.
It should be noted here that all the $Z_n$-flows in (\ref{hkp}) are
compatible, and this is a direct consequence of the definition
of $B_n$ \cite{sato:80}.  Namely we have:
\begin{lemma}
The differential operators $B_n$ in (\ref{hkp}) satisfy the following relation
called the Zakharov-Shabat equation,
\begin{equation}
\label{zs}
-i{\partial B_m \over \partial Z_n}+i{\partial B_n \over \partial Z_m}
+[B_m,B_n]=0,
\end{equation}
which gives the compatibility conditions among the flows in (\ref{lax}).
\end{lemma}
\begin{proof}
Consider the equation,
\begin{eqnarray*}
-i{\partial L^m \over \partial Z_n}+i{\partial L^n \over \partial Z_m}=
[B_n,L^m]-[B_m,L^n].
\end{eqnarray*}
Then splitting the operator $L^m$ into the parts having
nonnegative and negative powers of $D$, i.e. $L^m=B_m+B_m^C$ with $B_m^C:=
(L^m)_{<0}$, the right hand side can be written in the form,
\begin{eqnarray*}
[B_n,L^m]-[B_m,L^n]=[B_n,B_m+B_m^C]+[B_m^C,B_n+B_n^C].
\end{eqnarray*}
Projecting the equation on to the nonnegative (differential) part,
we obtain the result.
\end{proof}
\vspace{0.3cm}

Now we take the limit $ \be \to 0 $ which corresponds to a quasi-classical
limit of the NLS equation, and assume $ \phi $ to be of the WKB form,
\begin{equation}
\phi = \exp \left (i {1 \over \be} S \right ) \label{wkb}
\end{equation}
where $ S $ is called the action. Then, noting $D^{\nu}\phi/\phi
 \rightarrow (\partial S/\partial T)^{\nu}$ as $\beta \rightarrow 0$ for
 any integer $\nu$,  the eigenvalue problem $ L \phi = \la \phi $ in the
limit gives an algebraic equation for the spectral $\lambda$,
\begin{equation}
\la = P + {\rho \over {P - u}}, \label{genus0}
\end{equation}
where $P$ is the momentum defined by
\begin{equation}
P:={\partial S \over \partial T}=\lim_{\beta \to
0}\left(\frac{1}{\phi}D\phi\right).
\label{P}
\end{equation}
 The time evolution (\ref{kp}) with the rescaling $ Z \rightarrow Z / \be $
gives
\begin{equation}
{{\pat P} \over {\pat Z}} = {{\pat Q_2} \over {\pat T}}, \label{qeq}
\end{equation}
where $Q_n$ is defined by $\lim_{\beta\to 0}(B_n\phi/\phi)$,
which is just the polynomial part of $\lambda^n/n$ in $P$.
We denote this as $Q_n=[\la^n]_{\ge 0}/n$, and for example
$ Q_1 = [\la]_{\ge 0} =P,  Q_2 = [\la^2]_{\ge 0}  / 2=P^2/2+\rho $.
 Equation (\ref{qeq}) then provides the system (\ref
{g0}).  In particular, we note that the momentum $P$ gives the (polynomial)
conserved densities $P_n=P_n(\rho, u)$ of the system (\ref{g0}), i.e
\begin{equation}
P=\la -\sum^{\infty}_{n=1} \frac{P_n}{\la^n}=\la-\frac{\rho}{\la}-
\frac{\rho u}{\la^2} - \cdots.
\label{pexp}
\end{equation}
This is also true for the original NLS equation in the Lax form (\ref{lax}).
Namely, first express $D$ in the Laulent series of $L$ from (\ref{l}),
\begin{equation}
D=L-\sum_{n=1}^{\infty}D_nL^{-n}=L-\rho L^{-1}- \left(\rho u -\frac{1}{2}D\rho
\right)L^{-2} - \cdots,
\label{dexp}
\end{equation}
where $D_n$ are the differential polynomials of $\rho$ and $u$, which we denote
$D_n=D_n[\rho, u: \beta]$.  Also from Eq.(\ref{kp}) we have
\begin{equation}
-i{\partial \over \partial Z} \left(\frac{D\phi}{\phi}\right)=
D\left( \frac{B_2\phi}{\phi}\right)
\label{dcons}
\end{equation}
which shows that the $D_n$ are the conserved densities of the NLS equation.
The dispersionless limit of $D_n$ then gives $P_n$,
\begin{equation}
P_n(\rho, u) = \lim_{\beta \to 0} D_n[\rho, u:\beta].
\label{limcons}
\end{equation}
Note here that all the derivatives of $\rho$ and $u$ in $D_n$ vanish in
this limit.
This is valid if the functions $\rho$ and $u$ are both smooth in $T$.  However,
as seen in Fig.\ref{f2:4}, these functions develop high oscillations for
certain initial data.  Then one should extend the system (\ref{g0})
to the case including those oscillations.  This is indeed the main purpose
of this paper, and the extended system is called the Whitham equation
describing the averaged behaviour of the solution.

We also note from Eq.(\ref{hkp}) that the hierarchy can be reduced to
the following form with $Z_n\to Z_n/\beta$ and $Z_1=T$,
\begin{equation}
{{\pat Q_n} \over {\pat Z_m}} = {{\pat Q_m} \over {\pat Z_n}}, \label{hqeq}
\end{equation}
which give the compatibility conditions of the dispersionless limit of
(\ref{hkp}),
\begin{equation}
\label{qm}
{\partial S \over \partial Z_m}=Q_m.
\end{equation}
 Note that Eq.(\ref{qm}) with $P=\partial S/\partial T$ gives the
Hamilton-Jacobi
 equation with the hamiltonian $Q_m=Q_m(T,\partial S/\partial T, Z_n)$.
 This quasi-classical limit has been discussed in the framework of
 the dispersionless KP hierarchy (see for example
 \cite{aoyama:96,kodama:90}).

From this formulation, we now give a geometric interpretation
of the system (\ref{g0}).  First note that the algebraic equation (\ref
{genus0})
determines a two-sheeted Riemann surface of $ \la $ with branch points
$ \la_1 =u - 2 \sqrt {\rho} $ and $ \la_2 = u + 2 \sqrt {\rho} $.  Then
 after compactifying each sheet and gluing them together along the
branch cut, we have a Riemman surface of genus $ g = 0 $ which is defined by
the curve $y^2=R_0(\lambda)\equiv (\lambda-\lambda_1)
(\lambda-\lambda_2)$. Namely Eq.(\ref{genus0}) can be written
in the form $[P-(\la+u)/2]^2
=y^2/4$, and the roots are given by
\begin{equation}
\label{p}
P=\frac{1}{2}\left(\la +u \pm \sqrt{R_0(\la)}\right).
\end{equation}
  The branch points $\la_1$ and $\la_2$
are the Riemann invariants of the hyperbolic system and the values
$ P_1=P(\lambda_1)
$ and $ P_2 =P(\lambda_2)$ give the characteristic velocities which are also
given by the roots of the equation $\partial \la/\partial P=0$.
The system (\ref{g0}) then describes a (slow) modulation of the spectral
which is invariant for the NLS equation.  In Fig.\ref{f3:0},
we plot the curve (\ref{genus0}) and those points.
\begin{figure}
\epsfysize=7cm
\centerline{\epsffile{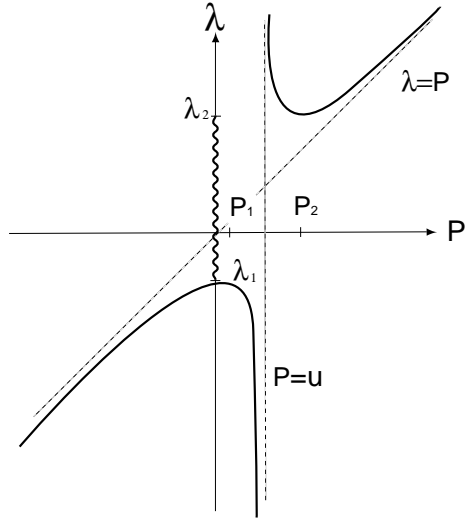}}
\caption{The $g=0$ algebraic curve (3.8).}
\end{figure}
 Note that the
spectral of $L$ is given by
Spec$(L)=(-\infty,\la_1]\cup [\la_2,\infty)$.  The $g=0$ Riemann surface
has a topology of a sphere as shown in Fig.\ref{f3:01}.  Thus
the hyperbolic system (\ref {g0}) can be
considered as an approximate system of the NLS equation on the $ g = 0 $
Riemann surface.
\begin{figure}
\epsfysize=6cm
\centerline{\epsffile{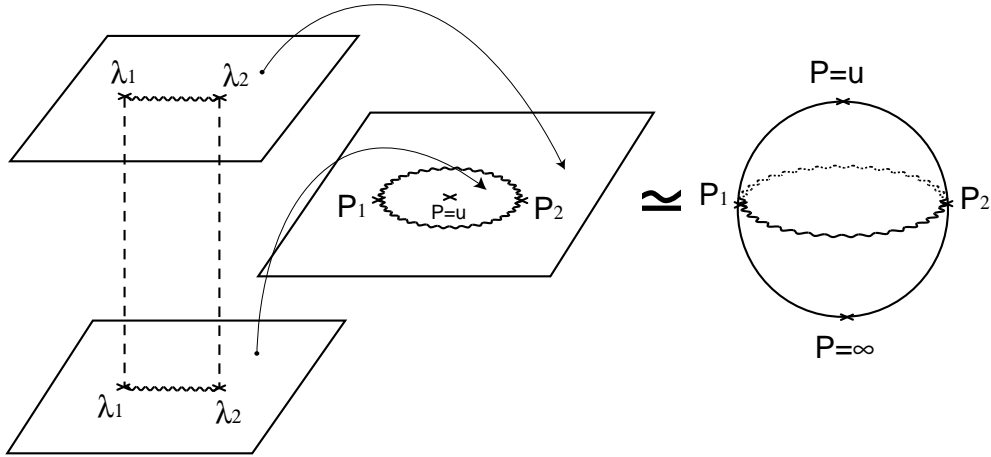}}
\caption{The $g=0$ Riemann surface corresponding to the curve (3.8).}
\end{figure}
Let us first demonstrate that the system (\ref{g0}) or (\ref{qeq}) can be
interpreted as a dynamics of the Riemann surface of genus 0.
On the $ g = 0 $ Riemann surface, we define the following meromorphic (Abelian)
differentials $ \om_1 $ and $ \om_2 $ of the second kind,
\begin{eqnarray}
& & \om_1 = {1 \over 2} \left(1 + {{\la - {1 \over 2} \sig_1} \over
{\sqrt {R_0}}} \right) d\la \sim d\lambda
+O\left(\frac{d\lambda}{\lambda^2}\right),
\label{diff1} \\
& & \om_2 ={1 \over 2} \left(\la + {{\la^2 - {1 \over 2} \sig_1 \la +
{1 \over 2} \sig_2-{1 \over 8} \sig_1^2} \over { \sqrt{R_0}}} \right)d\la
\nonumber \\
& & ~~~~~\sim \lambda d\lambda +O\left(\frac{d\lambda}{\lambda^2}\right),
\label{diff2}
\end{eqnarray}
\noindent
where $ \sig_1 = \la_1 + \la_2 (=2u)$ and $ \sig_2 = \la_1 \la_2 $. Note
from Eq.(\ref{p}) that these
differentials are given by $\omega_1=dQ_1=dP$ and $\omega_2=dQ_2=PdP$, that is,
$Q_1$ and $Q_2$ are the Abelian integrals.  In the case of $g=0$, these
differentials are trivial (image of the derivation), and it corresponds to
that the fundamental group for $g=0$ surface consists of only identity,
i.e.
for any cycle $a$ on the surface,
\begin{equation}
\oint_{a} \omega_i =0.
\label{trivial}
\end{equation}
Thus we have:
\begin{proposition}
\label{g0whitham}
The quasi-linear hyperbolic system (\ref{g0}) can be written by
\begin{equation}
{{\pat \om_1} \over {\pat Z}} = {{\pat \om_2} \over {\pat T}}, \label{whitham}
\end{equation}
where $\omega_1$ and $\omega_2$ are the Abelian differentials
(\ref{diff1}) and (\ref{diff2}) defined on the $g=0$ Riemann surface.
\end{proposition}
 Equation (\ref{whitham}) then defines the ($g=0$) Whitham equation
 \cite{flaschka:80,krichever:80}.  The differentials $\omega_1$ and
$\omega_2$ in Eq.(\ref
 {whitham}) are also given by the limits,
\begin{eqnarray}
& & \omega_1=dQ_1=d\left(\lim_{\beta\to 0}\frac{1}{\phi}D\phi\right),
\label{om1lim} \\
& & \omega_2=dQ_2=d\left(\lim_{\beta\to 0}\frac{1}{\phi}B_2\phi\right).
\label{om2lim}
\end{eqnarray}
Thus Eq.(\ref{whitham}) is just the compatibility condition of the
$T$- and $Z$-flows, that is, $\partial^2 \ln \phi/\partial T \partial Z
=\partial^2 \ln \phi/\partial Z \partial T$.
Cal\-cu\-lat\-ing the residue of
(\ref {whitham}) at each branch point $ \la_k $,
we obtain the Riemann invariant form of (\ref {whitham}),
\begin{equation}
{{\pat \la_k} \over {\pat Z}} = s_k (\la_1, \la_2) {{\pat \la_k}
\over {\pat T}}, ~~~~~ k = 1, 2, \label{ri0}
\end{equation}
where the characteristic velocity $s_k(\la_1, \la_2)$ is given by
\begin{equation}
s_k=\frac{\la^2-{1 \over 2}\sigma_1\la+{1 \over 2}\sigma_2 -{1 \over 8}
\sigma_1^2}{\la-{1 \over 2}\sigma_1}\Bigg|_{\la=\la_k}.
\label{skvelocity}
\end{equation}
Equation (\ref{ri0}) is of course the same as (\ref {rg0}).
Then we can rephrase the Proposition 2.1 as:
\begin{theorem}\label {thm1}
The $ g = 0 $ Whitham equation (\ref {ri0}) with the initial data (\ref
{l0}) has a unique global solution (rarefaction wave).
\end{theorem}

We now present the NLS-Whitham equations for arbitrary genus.
The two sheeted Riemann surface of genus $g$ is
defined by the algebraic (hyper-elliptic) curve $y^2=R_g(\lambda)$ with
\begin{equation}
R_g (\la) = \prod_{k=1}^{2 g + 2} (\la - \la_k), \label{genusg}
\end{equation}
where the branch points $ \la_k $ are real and are assumed to satisfy $
\la_1 < \la_2  < \ldots <  \la_{2g+2} $.  The Riemann surface of genus $g$
has a topology of a sphere with $g$ {\it handles}.  In Fig.\ref{f3:02}, we plot
the curve of $g=1$.  Comparing the case with $g=0$, we see that the closed curve
in the middle of the graph indicates an opening of the genus.
\begin{figure}
\epsfysize=7cm
\centerline{\epsffile{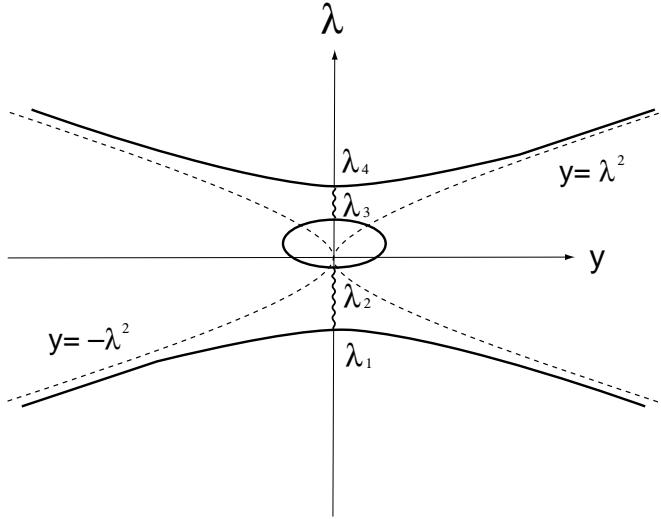}}
\caption{The $g=1$ algebraic curve $y^2=R_g(\la)$ in (\ref{genusg}).}
\end{figure}

 The NLS-Whitham equation on the genus $g$ Riemann surface is derived from
the NLS equation as the slow modulation of quasi-periodic wavetrain where
the
spectral of the Lax operator $L$ is given by,
 \begin{equation}
 \label{spec}
 \mbox{Spec} (L)={\bf R} \backslash \{\bigcup_{i=0}^g(\la_{2i+1},\la_{2i+2})\}.
 \end{equation}
 Here we reformulate the result in \cite{forest:88} as an extension of
 the $g=0$ NLS-Whitham equation (\ref{whitham}).
In the case of arbitrary genus $g$, the equivalent of the Abelian
differentials $\omega_1$ and $\omega_2$ defined before are given by
\begin{eqnarray}
& & \om_1 = {1 \over 2} \left(1 + {{\la^{g+1} - {1 \over 2} \sig_1 \la^g +
\alpha_1
\lambda^{g-1}+ \cdots +\alpha_g}
\over { \sqrt {R_g}}} \right)d \la  \nonumber \\
& & ~~~~~~ \sim d\la + O\left(\frac{d\la}{\la^2}\right), \ \mbox{as} \ \
\la \to \infty,
\label{diff1g}\\
& & \om_2 = {1 \over 2} \left( \la + {{\la^{g+2} - {1 \over 2}
\sig_1\la^{g+1} + (
{1 \over 2}\sigma_2-
{1 \over 8} \sig_1^2 )
\la^{g} +\beta_1\lambda^{g-1} + \cdots+ \beta_g} \over {\sqrt {R_g}}}
\right)d\la \nonumber \\
& & ~~~~~~ \sim \la d\la + O\left(\frac{d\la}{\la^2}\right),
\ \mbox{as} \ \ \la \to \infty,
\label{diff2g}
 \end{eqnarray}
where $\sigma_1$, $\sigma_2$ (and $\sigma_3$) are defined by
\begin{equation}
\sig_1 = \sum_{i=1}^4 \la_i, \ \ \sig_2 = \sum_{i < j} \la_i \la_j,
\ \ \sigma_3= \sum_{i<j<k} \la_i\la_j\la_k.
\label{sigma}
\end{equation}
The coefficients $\alpha_n$ and $\beta_n$ in (\ref{diff1g}) and (\ref{diff2g})
are determined by the normalizations on $\omega_1$ and $\omega_2$,
\begin{equation}
\oint_{a_i} \om_1 = \oint_{a_i} \om_2 = 0, \ \ \mbox{for} \ \ i=1,\cdots,g,
\label{normal}
\end{equation}
with the canonical $a_i$-cycle over the region $[\la_{2i+1},\la_{2i}]$.
Fig.ref{f3:1} shows the spectral of $L$ and the canonical cycles.
\begin{figure}
\epsfysize=3.5cm
\centerline{\epsffile{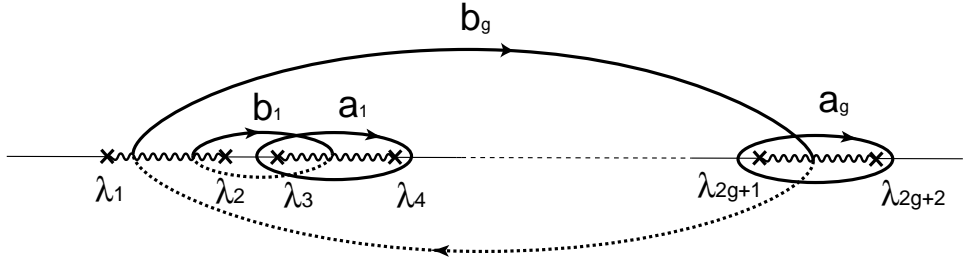}}
\caption{The 2-sheeted Riemann surface with branch cuts and canonical cycles.}
\end{figure}
Namely, we have the following $2g$ linear system of equations for
$\alpha_i$ and $\beta_i$ for $i=1,\cdots,g$,
\begin{eqnarray}
\left\{
\begin{array}{ll}
& \displaystyle{I_i^{g+1}-\frac{1}{2} \sigma_1 I_i^{g} +\alpha_1 I_i^{g-1}
+\cdots + \alpha_g I^0_i=0} , \\
& \displaystyle{I_i^{g+2}-\frac{1}{2} \sigma_1 I_i^{g+1}
+\left(\frac{1}{2}\sigma_2-\frac{1}{8}\sigma_1^2\right)I_i^{g}
+\beta_1 I_i^{g-1}+\cdots + \beta_g I_i^0=0},
\end{array}\right. \label{Ini}
\end{eqnarray}
where the integrals $I_i^n$ are given by
\begin{equation}
\label{int}
I_i^n=\frac{1}{2}\int_{\la_{2i+1}}^{\la_{2i+2}}\frac{\la^n}{\sqrt{-R_g}}d\la.
\end{equation}
Then the NLS-Whitham equation of genus $g$ is defined by the same form as
Eq.({\ref{whitham}).  As shown in \cite{forest:88,krichever:80,bloch:92},
the Whitham equation describes the conservation laws averaged over the high
oscillations appearing as a dispersive shock.
Then one can introduce the fast and slow variables to describe the oscillation
and the averaged behavior on the solution of the NLS equation
 in the weak dispersion limit \cite{whitham:78,flaschka:80}.
The average is then taken over the fast variable so that the averaged
equation describes the solution behaviour in the slow scale.  Thus the
average is given by
\begin{equation}
\label{average}
\langle f \rangle =\langle f \rangle(T_{slow}):=
\lim_{\beta\to 0}\frac{\beta}{2\pi}\int_{-\frac{\pi}{\beta}}^{
\frac{\pi}{\beta}}
f(T_{slow},T_{fast}) dT_{fast},
\end{equation}
where $T_{slow}$ and $T_{fast}$ are the slow and fast variables, and we use $T$
as the slow variable in the NLS-Whitham equation.  Note that the rescaling
$Z \to Z/\beta$ used to derive the Whitham equation (\ref{qeq})
defines the slow variable $Z=Z_{slow}$, i.e. $Z_{slow}=\beta Z_{fast}$.
 Then the Abelian
differentials $\omega_1$ and $\omega_2$ in the Whitham equation (\ref{whitham})
are defined by the averages \cite{bloch:92},
\begin{eqnarray}
& & \omega_1=d \Big{\langle}-i\beta\frac{\partial}{\partial T}\ln\phi
\Big{\rangle}\sim\left(1+\sum_{n=1}^{\infty}\frac{nF_n}{\la^{n+1}}\right)d\la,
\label{om1ave}\\
& & \omega_2=d \Big{\langle}-i\beta\frac{\partial}{\partial Z}\ln\phi
\Big{\rangle}\sim\left(\la+\sum_{n=1}^{\infty}
\frac{nG_n}{\la^{n+1}}\right)d\la.
\label{om2ave}
\end{eqnarray}
The Whitham equation leads to the conservation laws,
\begin{equation}
\label{c-law}
{\partial F_n \over \partial Z}={\partial G_n \over \partial T}
\end{equation}
in which the cases for $n=1$ and $n=2$ correspond to the averages of
(\ref{cons1}) and (\ref{cons2}).
Here $F_n$ and $G_n$ are the averaged conserved densities and fluxies,
and we have for example
\begin{eqnarray}
& & F_1=\left< \rho \right> = - {1 \over 4}\sigma_2 +{1 \over 16} \sigma_1^2
+ {1 \over 2}\alpha_1, \label{f1} \\
& & F_2=G_1 =\langle \rho u \rangle=
\frac{1}{4}\sigma_3-\frac{1}{8}\sigma_1\sigma_2
+\frac{1}{32} \sigma_1^3+\frac{1}{2}\beta_1 \nonumber \\
& & ~~~~~~~~~~~~~~~~~~~~~ =\frac{1}{8}\sigma_3-\frac{1}{8}\sigma_1\sigma_2
+\frac{1}{32}\sigma_1^3+\frac{1}{8}\sigma_1\alpha_1+\frac{1}{4}\alpha_2,
 \label{f2} \\
& & G_2= \left< \rho u^2 + {1\over 2}\rho^2 \right> =
-{ 1\over 8}\sigma_4 +{1\over 8}\sigma_1\sigma_3 + {1\over 32}\sigma_2^2-
{5 \over 64}\sigma_1^2\sigma_2 \nonumber \\
& & ~~~~~~~~~~~~~~~~~~~~~~~~~~~~ +{9 \over 512}\sigma_1^4
+{1 \over 8}\sigma_1\beta_1
+{1 \over 4}\beta_2.  \label{g2}
\end{eqnarray}
One should note that the quantities $F_n$ are given by the averages of $D_n$
in (\ref{dexp}), i.e. $F_n=\langle D_n \rangle$.
In the particular case of $g=0$,
we have $\alpha_i=\beta_i=0$, and $F_n$ become $P_n$ in (\ref{pexp}),
which means there is no oscillations in these quantities.

\section{Control of the NRZ pulses}
\renewcommand{\theequation}{4.\arabic{equation}}\setcounter{equation}{0}

We now consider the initial value problem of the NLS-Whitham equation with
the initial pulse having a nonzero chirp in a purpose of
reducing the pulse broadening due to nonlinearity and the dispersion.
The main objective is to determine the
average behavior of the solution of the NLS equation (\ref{nls}) in
the weak dispersion limit.  In a practical situation, this describes the
NRZ pulse behaviour through an optical filter which removes the high
oscillations (shock) due to the initial chirp.

\subsection{The hyperbolic structure of the NLS-Whitham
equation}

Let us first discuss the hyperbolic structure of the NLS-Whitham equation,
which enables us to analyze in detail the initial value problem.  As in the
case of
$g=0$, we have the Riemann invariant form by evaluating the residue at
$\la=\la_k$ in the NLS-Whitham equation (\ref{whitham}) with $\omega_1$
and $\omega_2$ given by (\ref{diff1g}) and (\ref{diff2g}),
\begin{equation}
\frac{\partial \la_k}{\partial Z}=s_k(\la_1,\cdots,\la_{2g+2})
\frac{\partial \la_k}{\partial T}, \ \ \mbox{for} \ k=1,\cdots,2g+2.
\label{rig}
\end{equation}
where the characteristic velocities $s_k$ is defined by the ratio
of the polynomials in $\omega_1$ and $\omega_2$, i.e.
\begin{equation}
s_k={{\la^{g+2} - {1 \over 2} \sig_1\la^{g+1} + (
{1 \over 2}\sigma_2-{1 \over 8} \sig_1^2 )
\la^{g} +\beta_1\lambda^{g-1} + \cdots+ \beta_g} \over
{\la^{g+1} - {1 \over 2} \sig_1 \la^g + \alpha_1
\lambda^{g-1}+ \cdots +\alpha_g}}\Bigg|_{\la=\la_k}. \label{velocity}
\end{equation}
Then following the ref.\cite{levermore:88}, we obtain:
\begin{lemma}
The characteristic velocities satisfy
\begin{eqnarray}
& & i) ~~~~~ {\partial s_{k} \over \partial \la_k} > 0, ~~~~~~~~ \
\mbox{for~} \ \
\forall k, \label{velo1} \\
& & ii) ~~~~~  s_{j}>s_{k}, ~~~~~~~~  \ \mbox{if~} \ \  \la_j>\la_k,
\label{velo2}
\end{eqnarray}
\end{lemma}
\begin{proof}
Let us call the polynomials in Eq.(\ref{velocity}) $\hat P(\la)$ for the
denominator and $\hat Q(\la)$ for the numerator.  Then from the normalizations
(\ref{normal}), we have
\begin{itemize}
\item
$\hat P(\la)$ has {\it exactly one} zero in the gap $[\la_{2i+1},\la_{2i+2}]$
for $i=0,1,\cdots,g$,
\item
$\hat Q(\la)$ has {\it at least one} zero in each gap.
\item
$s(\la):=\displaystyle{\frac{\hat Q(\la)}{\hat P(\la)}
 \to \lambda +O\left(\frac{1}{\la}\right)}$
as $\la \to \infty$.
\end{itemize}
From these proparties, the following function can be expressed as the curve in
Fig.\ref{f4:1};
\begin{equation}
\label{Vk}
V_{i}(\la):= s(\la) - s_{i} = \frac{1}{\hat P(\la)}
\left(\hat Q(\la) - s_{i}\hat P(\la)\right).
\end{equation}
It is then easy to see the relations (\ref{velo1}) and (\ref{velo2}).
\end{proof}
\vspace{0.3cm}

\begin{figure}
\epsfysize=7cm
\centerline{\epsffile{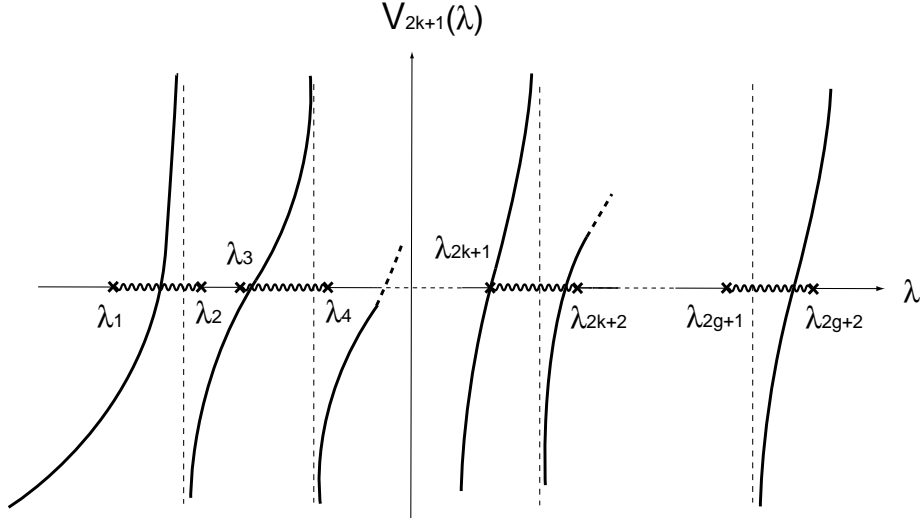}}
\caption{The function $V_i(\la)$ of (\ref{Vk}) with $i=2k+1$.}
\end{figure}

From this Lemma, we have the following useful Corollary which provides a way to
regularize the problem for the global existence of the solution:
\begin{corollary}
\label{coro1}
If all the initial $\la_k$ are monotonically decreasing and separated in
the sense that for $\forall T\in {\bf R}$ and $\forall k \in
\{1,2,\cdots,2g+2\}$
\begin{equation}
\label{order}
\min_{T}[\la_{k+1}(T,0)] > \la_k(T,0)>\max_T[\la_{k-1}(T,0)],
\end{equation}
then the NLS-Whitham
equation has a global solution.
\end{corollary}

\subsection{Regularization of shock singularity}

We now study the initial value problem of
the NLS-Whitham equation with the initial data shown in Fig.\ref{f4:1e}, i.e.
\begin{eqnarray}
& & \rho(T,0) =  \left \{\begin{array} {ll}
             \rho_0 , & \mbox{for~~~ $ |T| < T_0 $} \\
             0 , & \mbox{for~~~ $ |T| > T_0 $}
             \end{array} \right .  , \nonumber \\
  & & {} \label{initialg} \\
& & u(T,0) =  \left \{\begin{array} {lll}
             -u_0 ,  & \mbox{for~~~ $ -T_0< T < 0 $} \\
             u_0 ,& \mbox{for~~~ $ 0< T < T_0 $} \\
             0, & \mbox{otherwise}
             \end{array} \right . . \nonumber
\end{eqnarray}
\begin{figure}
\epsfysize=7cm
\centerline{\epsffile{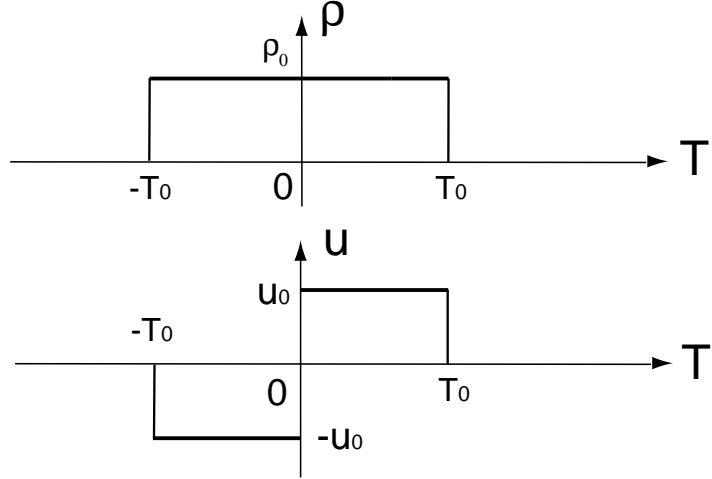}}
\caption{The initial phase modulation for one-bit NRZ pulse.}
\end{figure}
Here we vary the value of initial chirp $u_0$, and study its effect for a
purpose of reducing the NRZ pulse broadening.   As an example of real
situation,
we show in Fig.\ref{f4:2} a numerical result showing the effect of initial
chirp for a 16bit coded data (0010110010111100).  Here the initial phase
modulation is set to be periodic with the period given by the width of
one-bit pulse (bit-synchronous modulation), and the output
is smoothed out by using a filter (averaged over high oscillations).
As seen in the Figure, the overlapping of pulses in a) with no initial
phase is suppressed by the periodic initial phase modulation.  The period
is seen for example in (00111100) part as the elevation and depression of the
pulse level.  It looks that NRZ pulse with an initial phase modulation tends to
deform into an RZ pulse which may have a better property in signal processing
in a network system of optical communication.

Before we start to analyze this solution behaviour in detail, let us recall
that the value of the $u$ corresponds to the initial velocity of the
water, and the positive (negative) value of $u$ gives an compression
(expansion).  So we see a shock formation for the positive $u$ and a
rarefaction
wave for the negative $u$, which explains the pulse deformation shown
in Fig.\ref{f4:2}.  We also note from Eq.(\ref{l}) that the
characteristic speed for zero chirp is given by $2\sqrt{\rho}$, so that we
expect to see a different behavior of the solution for each $u_0$ value,
($i$) $u_0>2\sqrt{\rho_0}$,  ($ii$) $0<u_0<2\sqrt{\rho_0}$,
($iii$) $0>u_0>-2\sqrt{\rho_0}$, and ($iv$) $u_0<-2\sqrt{\rho_0}$.
This situation is quite similar to the case of Toda lattice equation discussed
in \cite{bloch:92}.  We then show that by choosing an appropriate number
of genus for a given initial data and by solving the corresponding
NLS-Whitham equation, one can obtain a global solution.  In the form
(\ref{initialg}), we here consider the case with $T_0=\infty$,
since the edges of the pulse with
constant chirp do not give any essential change from the case with zero chirp
except the velocity of the expansion.

\begin{figure}
\epsfysize=6.3cm
\centerline{\epsffile{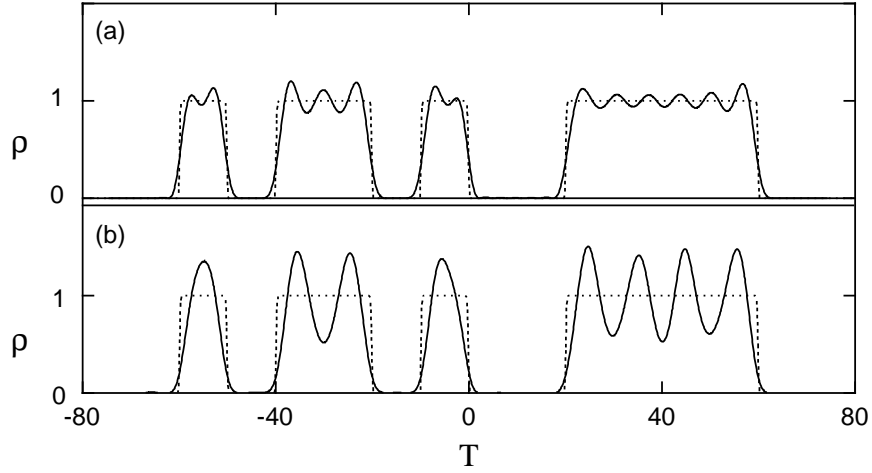}}
\caption{Effect of initial phase modulation for NRZ pulse propagation.
a) Without, and b) with the initial modulation.}
\end{figure}

We now consider each of the four cases separately:
\vspace{0.3cm}

\noindent
{\it (i) The case where $u_0>2\sqrt{\rho_0}$}:
\vspace{0.2cm}

In this case, the $g=0$ Riemann invariants $\la_1$ and $\la_2$ at $Z=0$ are
given by {\it increasing} functions, i.e.
\begin{eqnarray}
& & \la_1(T,0) =  \left \{\begin{array} {ll}
             -u_0-2\sqrt{\rho_0}, & \mbox{for~~~~ $ T < 0 $} \\
             u_0-2\sqrt{\rho_0} , & \mbox{for~~~~ $ T > 0 $}
             \end{array} \right .  ,\nonumber \\
 & & {} \label{initiali0} \\
& & \la_2(T,0) =  \left \{\begin{array} {ll}
             -u_0+2\sqrt{\rho_0} ,  & \mbox{for~~~~ $ T < 0 $} \\
             u_0 +2\sqrt{\rho_0} ,& \mbox{for~~~~ $ T >0 $}
             \end{array} \right . . \nonumber
\end{eqnarray}
Then we see from Corollary \ref{coro1}
 that the $g=0$ NLS-Whitham equation
develops a shock.  So we have to use the nonzero genus NLS-Whitham
equation for the regular solution.  First we note:
\begin{lemma}
\label{lemi}
The initial data can be identified as the following $g=1$ data,
\begin{eqnarray}
& & \la_1 = -u_0-2\sqrt{\rho_0}, ~~~~ \mbox{for~~ $\forall T$},  \nonumber \\
& & {} \nonumber \\
& & \la_2 =  \left \{\begin{array} {ll}
              -u_0+ 2\sqrt {\rho_0},  &  ~~~~ \mbox{for~~ $ T < 0$}, \\
              -u_0-2\sqrt{\rho_0}, &  ~~~~ \mbox{for~~ $ T > 0$}.
              \end{array} \right . \nonumber \\
& & {} \label{initiali} \\
& & \la_3 = \left \{\begin{array} {ll}
              u_0+2\sqrt{\rho_0}, &  ~~~~ \mbox{for~~ $ T < 0$}, \\
             u_0-2\sqrt{\rho_0},  &  ~~~~ \mbox{for~~ $ T > 0$},
             \end{array} \right .     \nonumber \\
   & & {} \nonumber \\
& & \la_4 = u_0+ 2 \sqrt {\rho_0}, ~~~~~ \mbox{for~~ $\forall T$},\nonumber
\end{eqnarray}
\end{lemma}
\vspace{0.3cm}
\begin{proof}
To prove the Lemma, we need to show the values $\langle \rho \rangle$
and $\langle \rho u \rangle$ in (\ref{f1}) and (\ref{f2}) for $g=1$ become
$\langle \rho \rangle=\rho$ and $\langle \rho u \rangle=\rho u$ at the
point $Z=0$, that is, the initial data is a {\it degenerate} $g=1$ data.
The quantities $\alpha_1$ and $\beta_1$ in (\ref{f1}) and (\ref{f2}) are
determined by the normalization
(\ref{normal}), that is, from (\ref{Ini})
\begin{eqnarray} \left\{
\begin{array}{ll}
& \displaystyle{I_1^2-\frac{1}{2} \sigma_1 I_1^1 +\alpha_1 I_1^0=0} , \\
& \displaystyle{I_1^3-\frac{1}{2} \sigma_1 I_1^2
+\left(\frac{1}{2}\sigma_2-\frac{1}{8}\sigma_1^2\right)I_1^1+\beta_1 I_1^0=0},
\end{array} \right. \label{alpha}
\end{eqnarray}
with
\begin{eqnarray*}
I_1^k =  \frac{1}{2}\int_{\la_3}^{\la_4} {{\la^k}
\over \sqrt {-(\la - \la_1)(\la - \la_2)(\la -
\la_3)(\la - \la_4)}} d\la.
\end{eqnarray*}
Substituting $\la=\la_3+(\la_4-\la_3)\sin^2\theta$ as in \cite{bloch:92},
 we have a convenient form of $I^k_1$,
\begin{displaymath}
I_1^k = \int_0^{\pi / 2} {{[\la_3 + (\la_4 - \la_3) \sin^2 \theta]^k}
\over \sqrt {[\la_3 - \la_1+(\la_4 - \la_3) \sin^2 \theta ] [\la_3 -
\la_2 + (\la_4 - \la_3) \sin^2 \theta]}} d\theta.
\end{displaymath}

Now let us compute $\langle \rho \rangle$ and $\langle \rho u \rangle$.
We first note that the Riemann invariants has the following symmetry in
the $T$-coordinate:
\begin{equation}
\label{lasym}
\la_k(-T)= -\la_{5-k}(T), \ \ \mbox{for} ~~~ k=1,\cdots,4,
\end{equation}
so that we have
\begin{eqnarray}
\begin{array}{lll}
& \displaystyle{ \sigma_k(-T)= (-1)^k\sigma_k(T),~~~~ \mbox{for} ~~~k=
1,\cdots,4} , \\
& {} \\
& \displaystyle{  I_1^n(-T)= (-1)^nI_1^n(T),
~~~~ \mbox{for} ~~~n=1,2,\cdots}.
\end{array} \label{sigsym}
\end{eqnarray}
Thus we consider only the case with $T<0$.  We then have
$I_1^k=\la_3^kI_1^0$ and
$\alpha_1=-u_0(u_0+2\sqrt{\rho_0}), \beta_1=2\rho_0(u_0+2\sqrt{\rho_0})$,
indicating that we {\it do} have $g=1$ initial data.  Using these values,
it is straightforward to show the result $\langle \rho \rangle=\rho_0$ and
$\langle \rho u \rangle=-\rho_0 u_0$ for $T<0$.
\end{proof}
\vspace{0.3cm}

\noindent
The point here is to recognize that the initial data is degenerate and
can also be identified as a non-zero genus data in which the Riemann invariants
are {\it decreasing} functions of $T$.
Then we have:
\begin{theorem}
\label{thmi}
For $u_0>2\sqrt{\rho_0}$, the $g=1$ NLS-Whitham equation with
the initial data (\ref{initiali}) has a global solution.
\end{theorem}
\begin{proof}
First note that $\la_k=$constant remains for $\forall Z$.  So we only need
to determine the evolutions for $\la_2$ and $\la_3$.  In order to compute
these values, we first regularize them at $T=0$:  From the relation
(\ref{velo2})
in Corollary \ref{coro1}, we have $s_2(T)<s_3(T)$ for $\forall T$, in
particular,
$s_2^-(0)<s_3^-(0)$
and $s_2^+(0)<s_3^+(0)$, where $s_k^{\pm}(0):=\lim_{\epsilon \to 0\pm}
s_k(\epsilon)$.  So if we impose
\begin{equation}
\label{limiti}
\la_2(T)\equiv \lim_{\epsilon \to 0+}\la_1(T-\epsilon), \ \ \ \mbox{and}
\ \ \ \la_3(T) \equiv \lim_{\epsilon\to 0-}\la_4(T-\epsilon),
\end{equation}
then we have $s_3^+>s_2^-$, and therefore from (\ref{velo1}) in Corollary
\ref{coro1}
we obtain the global solutions for $\la_2$ and $\la_3$.
We confirm this by caluculating explicitely
the evolution of the Riemann invariants.
The formula of $s_k:=s(\la)|_{\la=\la_k}$ for the $g=1$ case is given by
\begin{equation}
\label{sl}
s (\la) = {{\la^3 - {1 \over 2} \sig_1 \la^2 + ( {1 \over 2} \sig_2-{1
\over 8} \sig_1^2 ) \la + \beta_1} \over {\la^2 - {1 \over
2} \sig_1 \la + \alpha_1}},
\end{equation}
Using (\ref{alpha}), this can be written by
\begin{equation}
\label{sl1}
s(\la)=-\frac{1}{2}\sigma_1+{{\la^3 - {I_1^3 \over I_1^0}
+ ({1 \over
2} \sig_2-{3 \over 8} \sig_1^2 ) (\la - {I_1^1 \over I_1^0})} \over
{\la^2 - {I_1^2 \over I_1^0}
- {1 \over 2} \sig_1 (\la - {I_1^1 \over I_1^0})}}
\end{equation}

We now compute $s^-_3$ as well as $s^+_3, s^-_2$, and $s^+_2$.  However,
from the symmetry
discussed in the proof of Lemma \ref{lemi}, we have $s_3^-=-s_2^+,
s_3^+=-s_2^-$,
and also $s_3^->s_3^+$
from (\ref{velo1}) in Corollary \ref{coro1}.  So we need to compute only $s_3^-$
and $s_3^+$, which can be evaluated for the region of $T<0$.
In the computation of $s_3^-$,
we have $\la_1=-\la_3=-\la_4=-u_0-2\sqrt{\rho_0}$ and
$\la_2=-u_0+2\sqrt{\rho_0}$.
Then we find $I_1^k=\la_3^k I_1^0$, and obtain
\begin{equation}
\label{s3i}
s_3^-=(u_0+\sqrt{\rho_0})\left( 1+\frac{u_0\sqrt{\rho_0}}{(u_0+\sqrt{\rho_0})^2}
\right).
\end{equation}

In the case of $s_3^+$, we have $\la_1=-\la_4=-u_0-2\sqrt{\rho_0}$ and
$\la_2=-\la_3=-u_0+2\sqrt{\rho_0}$.  Then we obtain
\begin{equation}
\label{s2i}
s_3^+={u_0^2-6u_0\sqrt{\rho_0}+4\rho_0+6\sqrt{\rho_0}(u_0-2\sqrt{\rho_0})J_1
+8\sqrt{\rho_0}J_2 \over u_0 -2\sqrt{\rho_0}(1-J_1)}
\end{equation}
where $J_n, n=1,2,$ are given by
\begin{eqnarray*}
 J_n=\displaystyle{\int_0^{\pi/2}{\sin^{2n+2}\theta d\theta \over
\sqrt{(1+a^2\sin^2\theta)(1+a^2\cos^2\theta)}}\Big/ \int_0^{\pi/2}
{\sin^2\theta d\theta \over
\sqrt{(1+a^2\sin^2\theta)(1+a^2\cos^2\theta)}}} ,
\end{eqnarray*}
where $a^2=2\sqrt{\rho_0}/u_0<1$.
One can check the relation,
\begin{equation}
s_3^->s_3^+>0.
\label{s3relation}
\end{equation}
This proves that the $g=1$ NLS-Whitham equation with the initial
data (\ref{initiali}) admits a global solution.
\end{proof}
\begin{figure}
\epsfysize=6.5cm
\centerline{\epsffile{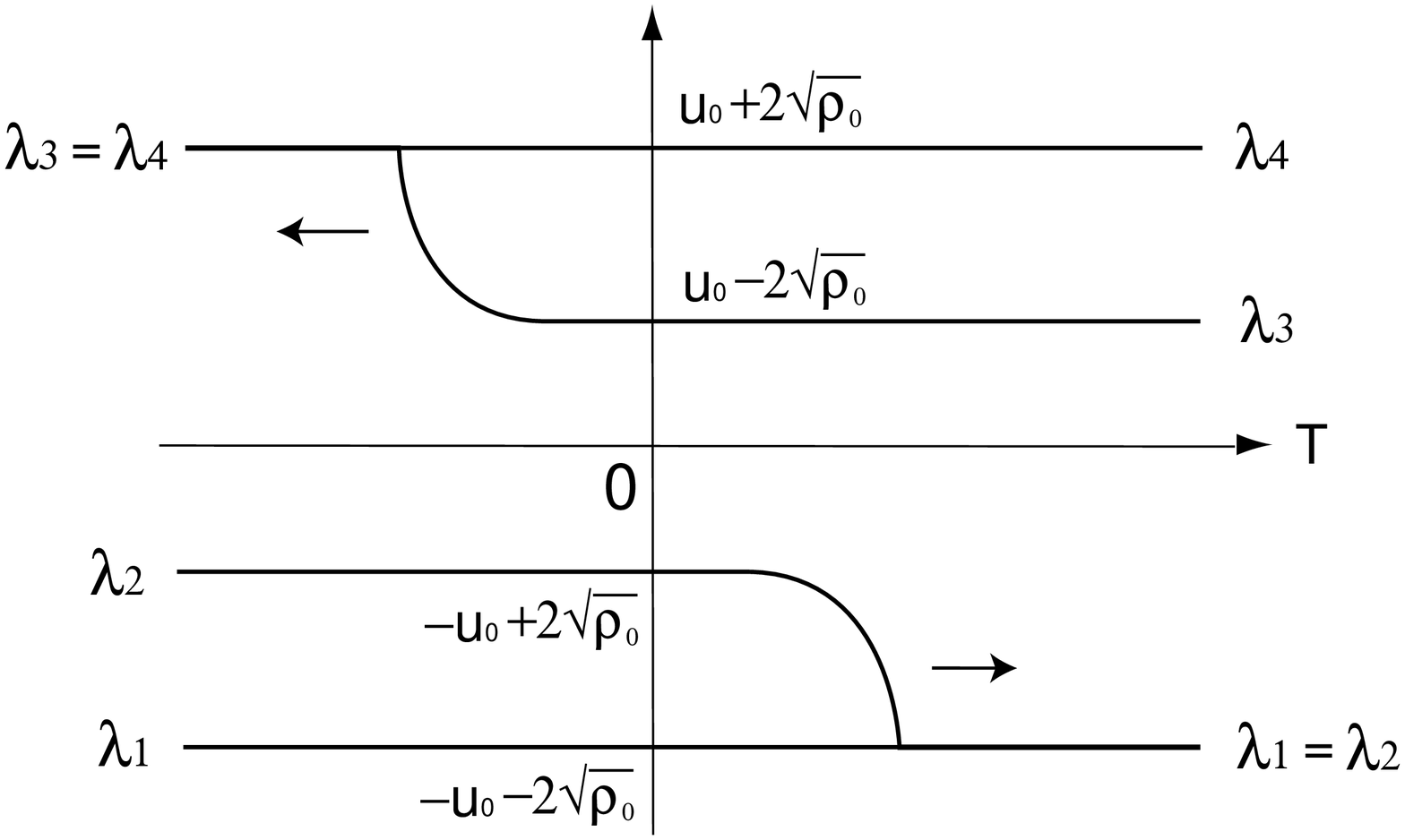}}
\caption{  Evolution of the Riemann invariants for $u_0>2 \sqrt{\rho_0}$. }
\end{figure}
\begin{figure}
\epsfysize=7cm
\centerline{\epsffile{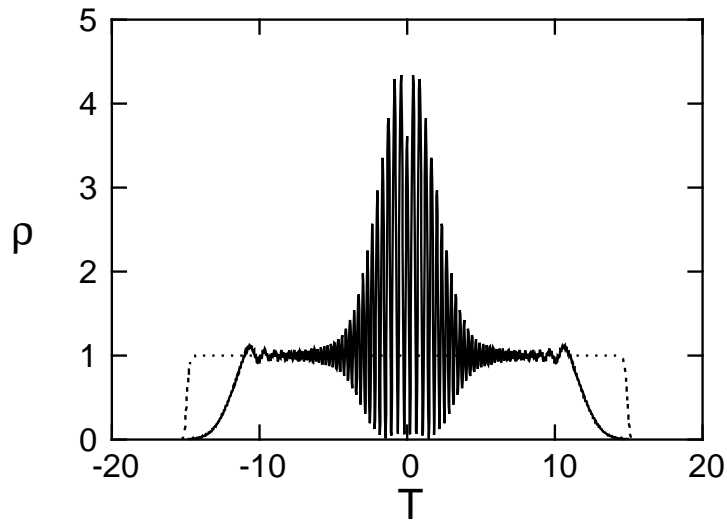}}
\caption{ Deformation of the pulse
$\rho(T,Z)$ for $u_0>2\sqrt{\rho_0}$. }
\end{figure}
\vspace{0.3cm}

\noindent
In Fig.\ref{f4:i1}, we plot the evolution of $\la_2$ and $\la_3$ together
with the
constants $\la_1$ and $\la_4$.  We also plot in Fig.\ref{f4:i2} the shape of
the pulse $ \rho(T,Z)$ at $Z=3\beta\approx 0.95$ obtained by the numerical
simulation of the NLS equation (\ref{nls}) with $\rho_0=1,
u_0=9\beta\approx 2.85$ and $\beta_2=-0.1$.  We see here nearly steady
oscillation in the center region $|T|<s_3^+Z$, which indicates the $g=1$
periodic
solution of the NLS equation as predicted in Fig.\ref{f4:i1}.
\vspace{0.3cm}

\noindent
{\it (ii) The case where $2\sqrt{\rho_0}>u_0>0$}:
\vspace{0.2cm}

This case also gives a compression of the pulse, and
we expect a shock formation in
the system Eq.(\ref{g0}).  So we again need to use the nonzero genus NLS-Whitham
equation for the regularization.  As in the previous case, we first
identify the initial
data (\ref{initialg}) as a non-zero genus data expressed as decreasing
functions:
\begin{lemma}
The initial data (\ref{initialg}) with $ 2\sqrt{\rho_0}>u_0>0$ can be considered
as a $g=2$ initial data,
\begin{eqnarray}
& & \la_1 = -u_0-2\sqrt{\rho_0}, ~ \ \mbox{for~ $\forall T$}, \nonumber \\
  & & {} \nonumber \\
& & \la_2 =  \left \{\begin{array} {ll}
              u_0-2 \sqrt {\rho_0},  &  \ \mbox{for} \ \ T < 0, \\
              -u_0-2\sqrt{\rho_0}, &  \ \mbox{for} \ \ T > 0.
              \end{array} \right . \nonumber \\
  & & {} \nonumber \\
& & \la_3 = u_0-2\sqrt{\rho_0}, ~ \ \mbox{for~ $\forall T$}, \nonumber \\
 & & {} \label{initialg2} \\
& & \la_4 = -u_0 + 2\sqrt{\rho_0} ,~ \ \mbox{for~ $\forall T$}, \nonumber \\
 & & {} \nonumber \\
& & \la_5 = \left \{\begin{array} {ll}
              u_0+2\sqrt{\rho_0}, &  \ \mbox{for} \ \ T < 0, \\
             -u_0+2\sqrt{\rho_0},  &  \ \mbox{for} \ \ T > 0,
             \end{array} \right .     \nonumber \\
  & & {} \nonumber \\
& & \la_6 = u_0+ 2 \sqrt {\rho_0}, ~ \ \mbox{for~ $\forall T$},\nonumber
\end{eqnarray}
\end{lemma}
\begin{proof}
The reflective symmetry in the data mentioned in the previous case
 holds here as well, and we have
\begin{equation}
\label{symii}
\la_k(-T)= -\la_{7-k}(T), ~~~~ \ \mbox{for ~~~} k=1,\cdots,6,
\end{equation}
which lead to
\begin{equation}
\label{sigmaii}
\sigma_k(-T)= (-1)^k\sigma_k(T), ~~~~ \mbox{for} ~~~ k=1,\cdots,6.
\end{equation}
The integrals $I_i^n$ of (\ref{int}) with $g=2$ also satisfy
\begin{equation}
\label{Iii}
I_i^n(-T)= (-1)^nI_i^n(T), ~~~~ \mbox{for} ~~~ i=1,2, ~~\mbox{and ~~}
n=1,2,\cdots.
\end{equation}
As in the previous case, we need to compute $\langle \rho \rangle$ and
$\langle \rho u \rangle$ for only $T<0$.

  From (\ref{Ini}), we see for $i=0,1$
\begin{eqnarray}
& &
\displaystyle{\alpha_{i+1}=\frac{(-1)^{i+1}}{[1,0]}\left([3,i]-\frac{1}{2}
\sigma_1
[2,i]\right)}, \label{alphai} \\
& & \displaystyle{\beta_{i+1}=\frac{(-1)^{i+1}}{[1,0]}\left\{
[4,i]-\frac{1}{2}\sigma_1
[3,i]+\left(\frac{1}{2}\sigma_2-\frac{1}{8}\sigma_1^2\right)[2,i]\right\}},
\label{betai}
\end{eqnarray}
where
\begin{eqnarray*}
[n,m]:=\det\left(\matrix {I_1^n & I_1^m \cr I_2^n & I_2^m \cr}\right) .
\end{eqnarray*}
For $T<0$ we note that $\la_1=-\la_5=-\la_6$ and $\la_2=\la_3=-\la_4$.
Then caluculating $[n,m]$ leads to
$\alpha_1=-(\la_1^2+\la_2^2)/2, \beta_1=(\la_2-\la_1)^3/8$, and using
$\sigma_1=(\la_2-\la_1), \sigma_2=-(\la_1^2+\la_2^2+\la_1\la_2),
\sigma_3=(\la_1-\la_2)(\la_1^2+\la_2^2)$, we verify
\begin{eqnarray}
& & \displaystyle{\langle \rho \rangle=\frac{1}{2}(\la_1+\la_2)^2=\rho_0} ,
\label{rhoii} \\
& & \displaystyle{\langle \rho u
\rangle=\frac{1}{32}(\la_1+\la_2)^2(\la_1-\la_2)
=-\rho_0 u_0}.  \label{rhouii}
\end{eqnarray}
This completes the proof.
\end{proof}
\vspace{0.3cm}
From this lemma, we obtain:
\begin{theorem}
\label{thm2}
For $ 2\sqrt{\rho_0}>u_0>0$, the $g=2$ NLS-Whitham equation with the initial
data (\ref{initialg2}) admits the global solution.
\end{theorem}
\begin{proof}
We need to compute the velocities $s^-_5(=-s_2^+)$ and $s^+_5(=-s_2^-)$.
The regularization of the initial data is the same as the previous case, and
we here impose
\begin{equation}
\label{regii}
\la_2(T)\equiv \lim_{\epsilon \to 0+}\la_2(T-\epsilon), ~~~~
\la_5(T)\equiv \lim_{\epsilon \to 0-}\la_5(T-\epsilon).
\end{equation}
Then calculating the integrals given by the normalizations (\ref{normal}), we
obtain the follwing results,
\begin{equation}
\begin{array}{rcl}
& & \displaystyle{s_5^-=-s_2^+=(u_0+\sqrt{\rho_0})
\left(1+\frac{u_0\sqrt{\rho_0}}{( u_0+\sqrt{\rho_0})^2}\right)}, \\
& & \displaystyle{s_5^+=-s_2^-= -\frac{1}{2} u_0+\sqrt{\rho_0}}>0.
\end{array}
 \label{sii}
\end{equation}
(We omit the details of the calculation, since
the procedure is the same as in \cite{bloch:92} and is straightforward but
so tedious.)
This confirms the existence of the global solution.
\end{proof}
\vspace{0.3cm}
The solution $\langle \rho \rangle$ can be caluculated from
 (\ref{f1}), and in the region $|T|<s_5^+Z$, it takes the constant value,
\begin{equation}
\label{irho}
\langle \rho \rangle =\rho_0\left(1+\frac{u_0}{2\sqrt{\rho_0}}\right)^2>\rho_0.
\end{equation}
This explains the elevation of the pulse level observed in Fig.\ref{f4:2}.
In Fig.\ref{f4:ii}, we plot the evolution of the Riemann invariants.  The
corresponding solution $\rho(T,Z)$ at $Z=5\beta\approx 1.58$ of the
numerical simulation of the NLS equation (\ref{nls}) is
shown in Fig.\ref{f2:4}, where $\rho_0=1, u_0=3\beta \approx 0.95$ with
$\beta_2=-0.1$.
Note in Fig.\ref{f2:4} that the non-oscillating part with the
constant level (\ref{irho}) in the center region $|T|
<s_5^+Z$ corresponds to the $g=2$ (degenerated into $g=0$) solution, and
the oscillating parts in the regions $s_5^+Z<|T|<s_5^-Z$ to the ones
(degenertated into $g=1$), as predicted in Fig.\ref{f4:ii}.  One should
compare Fig.\ref{f2:4} with Fig.\ref{f4:i2} to see the difference in the
center
regions.
\vspace{0.3cm}

\begin{figure}
\epsfysize=6.5cm
\centerline{\epsffile{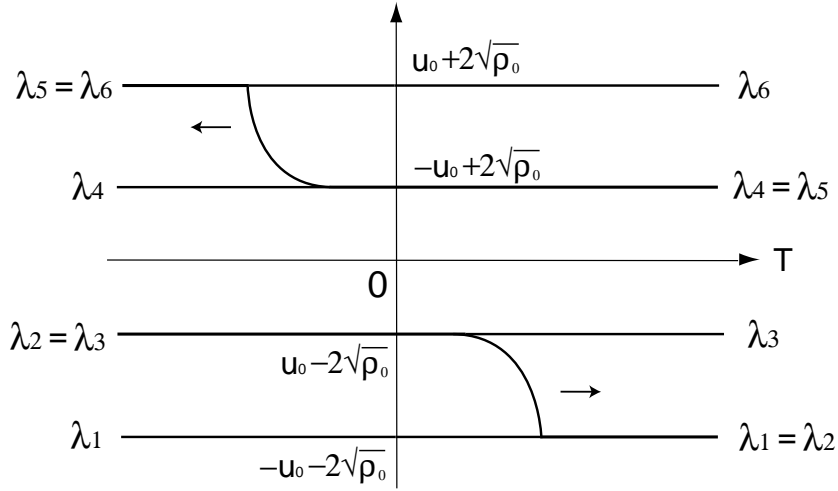}}
\caption{ Evolution of the Riemann invariants for $0<u_0<2 \sqrt { \rho_0 }$. }
\end{figure}

\noindent
{\it (iii) The case with $0>u_0>-2\sqrt{\rho_0}$}:
\vspace{0.2cm}

In this case, the initial data gives decreasing functions for
both $\la_1$ and $\la_2$ (see Fig.\ref{f2:2}), and we have:
\begin{theorem}
\label{thm3}
For $ 0>u_0>-2\sqrt{\rho_0}$, the $g=0$ NLS-Whitham equation admits the global
solution with the initial data (\ref{initialg}).
\end{theorem}
\begin{proof}
In this case, $\la_1$ and $\la_2$ can be regularized as
\begin{equation}
\label{regla12}
\la_1(T)\equiv \lim_{\epsilon \to 0+}\la_1(T-\epsilon), \ \ \ \mbox{and}
\ \ \ \la_2(T) \equiv \lim_{\epsilon \to 0-} \la_2(T-\epsilon).
\end{equation}
Then from Eq.(\ref{velocity}) of $s_k$ for $g=0$, we obtain
\begin{equation}
\begin{array}{rcl}
& & \displaystyle{s_2^-=-s_1^+= -u_0+\sqrt{\rho_0}>0},  \\
& & \displaystyle{s_2^+=-s_1^-= \frac{1}{2} (u_0+2\sqrt{\rho_0}) >0}.
\end{array}
 \label{s2}
\end{equation}
Thus we have $s_2^->s_2^+>0>s_1^->s_1^+$, which implies the assertion of
the theorem.
\end{proof}
\vspace{0.3cm}
\begin{figure}
\epsfysize=6.5cm
\centerline{\epsffile{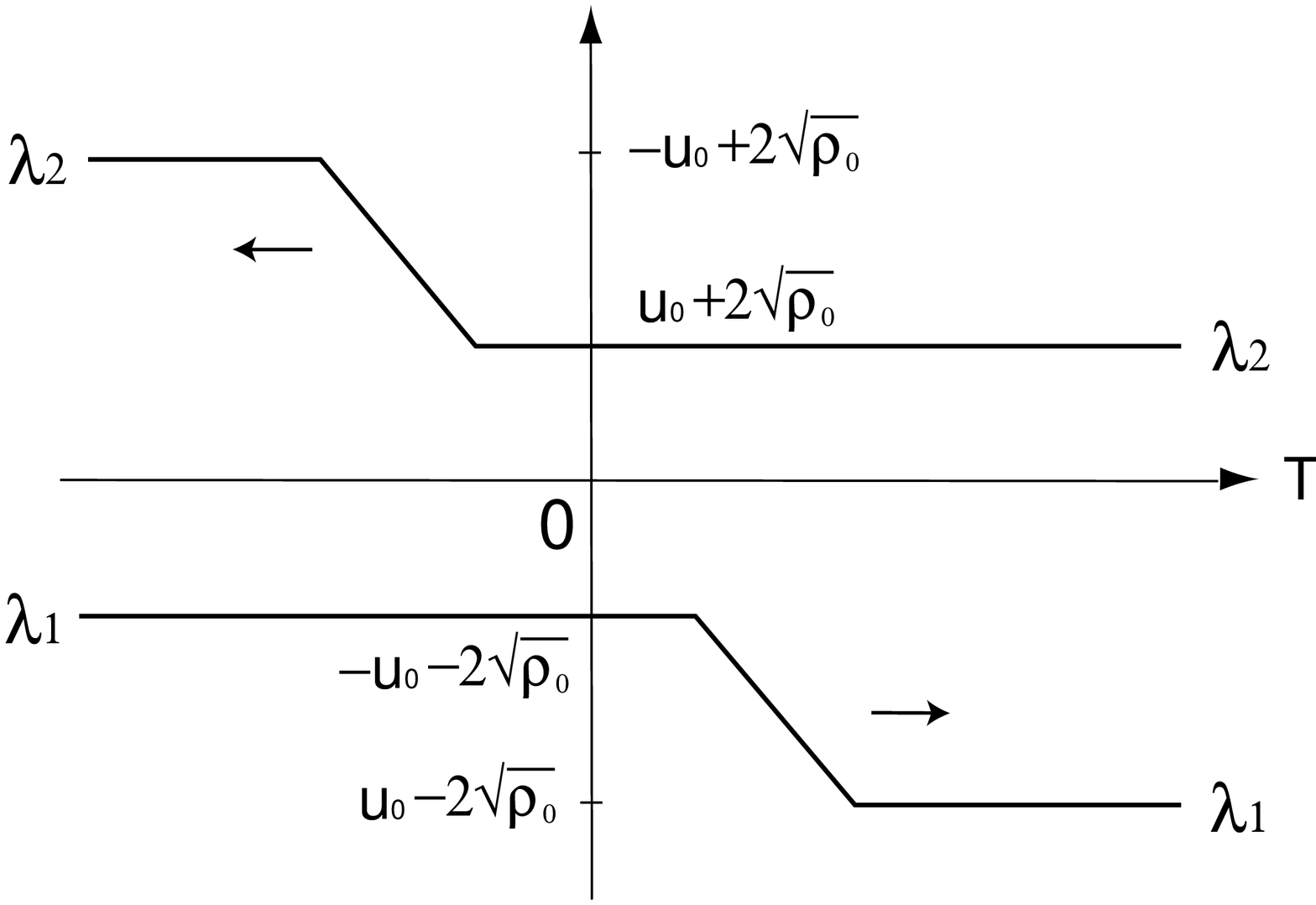}}
\caption{ Evolution of the Riemann invariants for $0>u_0>-2\sqrt{\rho_0}$.}
\end{figure}
\begin{figure}
\epsfysize=6cm
\centerline{\epsffile{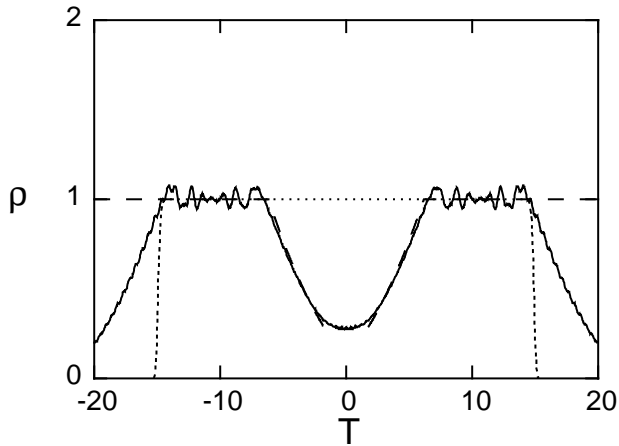}}
\caption{ Deformation of the pulse $\rho(T,Z)$ for $0>u_0>-2\sqrt{\rho_0}$:
Solid curve, the numerical result;
Dashed curve, the solution of Eq.(\ref{rhoiii}). }
\end{figure}

\noindent
We also find the solutions $\la_1, \la_2$ and $\langle
\rho(T,Z)\rangle=\rho(T,Z)$:
For $T>0$,
\begin{eqnarray}
& & \la_1(T,Z)=
\left \{\begin{array} {lll}
      \displaystyle{u_0-2 \sqrt{\rho_0}},
       & \mbox{for~~~  $ T >s_2^{-}Z $}, \\
      \displaystyle{ -\frac{4}{3Z} T - \frac{1}{3}(u_0+2\sqrt{\rho_0})},
      & \mbox{for~ $ s_2^{+}Z< T < s_2^{-}Z  $}, \\
      \displaystyle{ -u_0-2\sqrt{\rho_0}},
     & \mbox{for~~~ $ 0<T <s_2^{+}Z $},
             \end{array} \right . \nonumber \\
     & & {} \label{la12} \\
& & \la_2(T,Z)=u_0+2\sqrt{\rho_0}, ~~~~~~ \mbox{for~~~ $\forall T>0$,} \nonumber
\end{eqnarray}
and
\begin{equation}
\rho(T,Z)=
\left \{\begin{array} {lll}
      \displaystyle{{\rho_0}},
       & \mbox{for~~~ $ T >s_2^{-}Z $}, \\
     \displaystyle{ \frac{1}{9Z^2} \left\{T+(u_0+2\sqrt{\rho_0})Z
     \right\}^2},
      & \mbox{for~~ $ s_2^{+}Z< T < s_2^{-}Z $}, \\
     \displaystyle{\rho_0\left(1+\frac{u_0}{2\sqrt{\rho_0}}\right)^2},
     & \mbox{for~~~ $ 0<T <s_2^{+}Z $},
             \end{array} \right . \label{rhoiii}
\end{equation}
For $T<0$, note the symmetry $\la_1(-T,Z)=-\la_2(T,Z)$ and
$\rho(-T,Z)=\rho(T,Z)$.
These solutions are shown in Fig.\ref{f4:iii1} and Fig.\ref{f4:iii2}.
In Fig.\ref{f4:iii2}, we also plot the numerical result $\rho(T,Z)$ at
$Z=10\beta\approx 3.16$ of the NLS equation (\ref{nls}) where $\rho_0=1,
u_0=-3\beta\approx -0.95, \beta_2=-0.1$, which agrees quite well with the
analytical solution (\ref{rhoiii}).  Note that the level of $\rho$ for
$|T|<s_2^+Z$ has the same formula as the previous case except the sign of
$u_0$.
This explains the bit-wise depression in the pulse level shown in
Fig.\ref{f4:2}, and it leads to the deformation into a RZ-like
pulse observed in the experiment \cite{bergano:96}.  This also gives a
theoretical limit of the strength of the initial chirp, i.e.
$u_0>-2\sqrt{\rho_0}$, to avoid a destruction of the pulse.
In fact, we will see below that the case
with $u_0<-2\sqrt{\rho_0}$ the level of the signal becomes zero
starting from $T=0$, the point of the discontinuity in chirp.
\vspace{0.3cm}

\noindent
{\it (iv) The case where $u_0<-2\sqrt{\rho_0}$}:
\vspace{0.2cm}

  We first have:
\begin{lemma}
The initial data (\ref{initialg}) for $u_0<-2\sqrt{\rho_0}$
can be considered as a
$g=1$ initial data where the Riemann invariants $\la_1,\cdots,\la_4$ have
the following values,
\begin{eqnarray}
& & \la_1 = \left \{\begin{array} {ll}
              u_0+2\sqrt{\rho_0},  & \mbox{for~~ $ T < 0 $,} \\
             u_0-2\sqrt{\rho_0},  & \mbox{for~~ $ T >  0 $,}
             \end{array} \right . \nonumber \\
    & & {} \nonumber \\
& & \la_2 = u_0+ 2 \sqrt {\rho_0}, ~~ \mbox{for~~$\forall T$,}\nonumber \\
& & {} \label{initialiv} \\
& & \la_3 = -u_0-2\sqrt{\rho_0},~~ \mbox{for~~$\forall T$,} \nonumber \\
    & & {} \nonumber \\
& & \la_4 =  \left \{\begin{array} {ll}
              -u_0+2 \sqrt {\rho_0},  & \mbox{for~~ $ T < 0 $,} \\
              -u_0-2\sqrt{\rho_0}, & \mbox{for~~ $ T > 0 $.}
             \end{array} \right .\nonumber
\end{eqnarray}
\end{lemma}
\begin{proof}
The calculation to show $\langle \rho \rangle=\rho$ and $\langle \rho u \rangle
=\rho u$ is the same as the previous cases, and we omit it.
\end{proof}
\vspace{0.3cm}

\begin{figure}
\epsfysize=6.5cm
\centerline{\epsffile{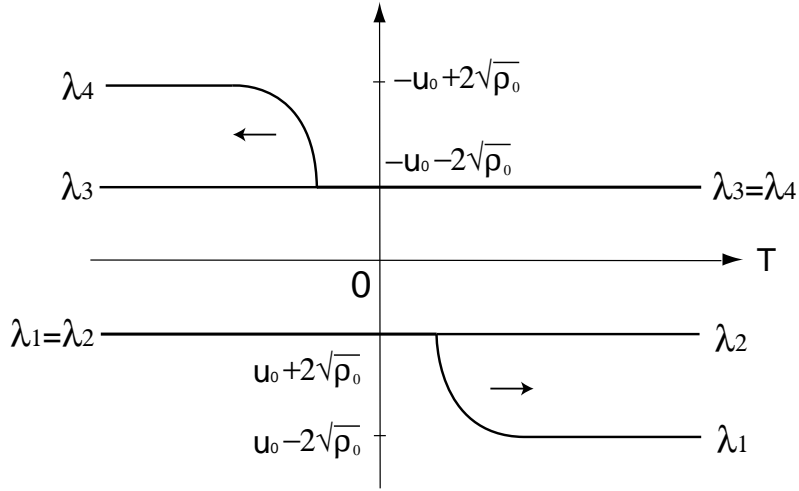}}
\caption{ Evolution of the Riemann invariants for $u_0<-2\sqrt{\rho_0}$.}
\end{figure}

\noindent
In Fig.\ref{f4:iv1}, we plot the initial data as the solid lines.
Since all the Riemann
invariants $\la_k$ satisfy the conditions in Corollary \ref{coro1},
we obtain:
\begin{theorem}
\label{thmg1}
For $u_0<-2\sqrt{\rho_0}$, the $g=1$ NLS-Whitham equation with the initial
data (\ref{initialiv}) has a global solution.
\end{theorem}
\begin{proof}
We determine the evolutions for $\la_1$ and $\la_4$.  The regularization of
$\la_1$ and $\la_4$ at $T=0$ is given by imposing
\begin{equation}
\label{limitl}
\la_1(T)\equiv \lim_{\epsilon \to 0+}\la_1(T-\epsilon), \ \ \ \mbox{and}
\ \ \ \la_4(T) \equiv \lim_{\epsilon\to 0-}\la_4(T-\epsilon).
\end{equation}
Then we have $s_4^+>s_1^-$, and therefore we obtain the global solution as
in the previous cases.
In fact, from the formula of $s_k$, we have
\begin{equation}
\begin{array}{rcl}
& & \displaystyle{s_4^-=-s_1^+= -u_0+\sqrt{\rho_0}>0},  \\
   & & {} \\
& & \displaystyle{s_4^+=-s_1^-= - u_0-2\sqrt{\rho_0}>0}.
\end{array}
 \label{siv}
\end{equation}
This completes the proof.
\end{proof}
\vspace{0.3cm}

\noindent
We note in particular that the solution $\langle \rho(T,Z) \rangle$ in the
center region $|T|<s_4^+Z$ becomes zero, which is obtained from
Eq.(\ref{f1}) with $\sigma_1=0, \sigma_2=-2\la_3^2, I_1^k=\la_3^k I_1^0$
and $\alpha_1=-\la_3^2$. In Fig.\ref{f4:iv2}, we plot the pulse profile
$\rho(T,Z)$ at $Z=3\beta\approx 0.95$ obtained by the numerical simulation
of the NLS equation (\ref{nls}) with $\rho_0=1, u_0=-9\beta\approx -2.85$
and $\beta_2=-0.1$.  As was predicted by the theory, the pulse level at
$T=0$
becomes zero right after
the propagation.  One should also note that the genus is actually degenerated
into $g=0$ for $\forall T$, even though we need the $g=1$ regularized initial
data for the global solution.  This implies that the solution is indeed
$g=1$, but there is no oscillations in the solution.  The small oscillations
observed in Fig.\ref{f4:iv2} are due to a nonzero value of $\beta_2$, and
disappear in the limit $\beta_2 \to 0$.
\begin{figure}
\epsfysize=7cm
\centerline{\epsffile{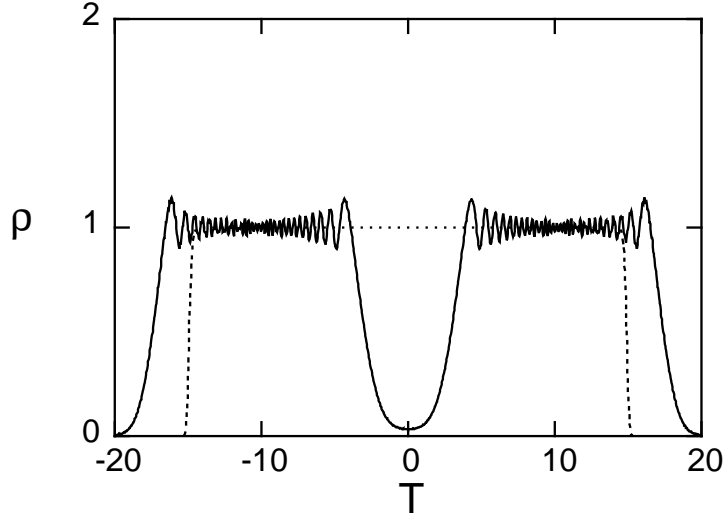}}
\caption{Deformation of the pulse $\rho(T,Z)$ for $u_0<-2\sqrt{\rho_0}$.}
\end{figure}

In this Section, we have considered the evolution of the Riemann invariants
for $Z>0$ in the cases with several different values of initial chirp
$u_0$.
We then recognized that some
of the cases gave {\it degenerate} genus initial data, and choosing appropriate
genus, we found the global solutions for the corresponding NLS-Whitham
equations.
We also note that the original system of equations (\ref{cons1}) and
(\ref{cons2})
is invariant under the change of variables $Z\to -Z$ and $u\to -u$.  Hence
for example the evolution of the case ($ii$) for $Z>0$ can be extended
to the evolution for $Z<0$ with the case ($iii$) under the change
$Z\to -Z$ and $u_0\to -u_0$.  Thus we have described a change of genus along
the $Z$-axis, and in the example we see the Riemann surface of $g=0$ for $Z<0$
and of $g=2$ (degenerated into $g=1$) for $Z>0$.  This process is illustrated
in Fig.\ref{f4:5}.

\begin{figure}
\epsfysize=4cm
\centerline{\epsffile{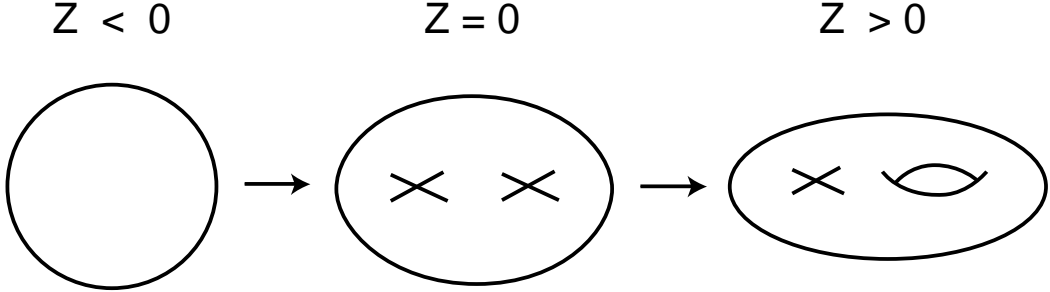}}
\caption{Deformation of the Riemann surface for {$0<u_0<2\sqrt{\rho_0}$}.}
\end{figure}

\vspace{0.3cm}

\section{Effects of the higher order dispersion}
\renewcommand{\theequation}{5.\arabic{equation}}\setcounter{equation}{0}

As shown in \cite{kodama:85,kodama:87}, the important higher order effects
in a long-distance and high-speed optical communication system are
described by the perturbed NLS equation,
\begin{equation}
\label{hnls}
i{\partial q \over \partial Z} + {\beta_2 \over 2}{\partial^2 q \over
\partial T^2}
+ |q|^2q = i {\beta_3 \over 6}{\partial^3  q \over \partial T^3}
+ i \gamma {\partial \over \partial T}(|q|^2q) + \sigma_R q{\partial \over
\partial T}
|q|^2,
\end{equation}
where $\beta_3, \gamma, \sigma_R$ are real constants describing the third order
linear dispersion, nonlinear dispersion due to the delay effect of the Kerr
nonlinearity and the Raman (dissipative) effect.  Since the NRZ system
operates in a small GVD ($\beta_2$) regime, the third order dispersion
($\beta_3$) becomes particularly important.
Also the Raman term may not be so important in the
regime of pulsewidth with scores of pico-second.  So we consider the
effects of $\beta_3$ and $\gamma$ in Eq.(\ref{hnls}).  In a real system, we
also use a region where $\beta_3$  becomes (or make)
 so small, and assume that $\beta_3\propto (\beta_2)^{3/2}$.  Then as
 shown in \cite{kodama:85} Eq.(\ref{hnls}) with $\sigma_R=0$ can be
Lie-transformed to an integrable system (the NLS + its hierarchy),
\begin{equation}
\label{hnls3}
i{\partial q \over \partial Z} + {\beta_2 \over 2}{\partial^2 q \over
\partial T^2}
+  |q|^2q = i {\beta_3 \over 6}{\partial^3  q \over \partial T^3}
+ i {\beta_3 \over \beta_2}|q|^2 {\partial q  \over \partial T}+
(\mbox{higher~orders}).
\end{equation}
In a small $\beta_2$, the transformation is however valid only for a weak
nonlinear case, e.g. $|q|^2 \sim O(\beta_2)$.
Equation (\ref{hnls3}) without the higher orders has also infinite number
of conservation laws with the same
conserved densities as the case of the NLS equation.  In fact, this can be
also written in the Lax form,
\begin{equation}
-i{\partial L \over \partial Z}=[B, L], \ \ \mbox{with} \ \
B=B_2+ \hat \beta B_3, \label{tod}
\end{equation}
where $B_n$ are defined in (\ref{hkp}), and
$\hat \beta =-\beta_3/(2\beta^3)$ with $\beta_2\equiv -\beta^2$.  The
explicit formula
of $B_3$ is given by
\begin{equation}
\label{b3}
B_3=\frac{1}{3} D^3 +\rho D + \rho u + \frac{1}{2} D \rho .
\end{equation}

Now let us take the limit $\beta \to 0$, and assume both $\rho$ and $u$ are
smooth.  Then as in the previous section we have the $g=0$ approximation of
the equation (\ref{hnls3}), which is obtained in the form similar to
(\ref{qeq}) with the rescaling $Z \to Z/\beta$,
\begin{equation}
\label{qeq3}
{\partial Q_1\over \partial Z}={\partial \over \partial T}(Q_2+\hat \beta Q_3),
\end{equation}
where $Q_3=[\la^3]_{\ge 0}=P^3/3+\rho P + \rho u$.  In terms of $\rho$ and $u$,
Eq.(\ref{qeq3}) gives a system,
\begin{eqnarray}
{\pat \over {\pat Z}} \left(\matrix {\rho \cr  u \cr}
\right) & = & \left(\matrix { u+\hat \beta (u^2+2\rho) &
\rho + \hat\beta (2u\rho) \cr  1 +\hat\beta (2u) &  u+\hat\beta(u^2+2\rho)
\cr} \right) {\pat \over {\pat T}} \left(\matrix {\rho \cr
u \cr} \right).  \label{g0t}
\end{eqnarray}
Since the Lax operator for the equation (\ref{hnls3}) is  the same as
the NLS case, the spectral curve in this limit is  given by
(\ref{genus0}), and the system (\ref{g0t}) is also referred as the
$g=0$ Whitham equation.
The eigenvalues of the coefficient matrix of (\ref{g0t}) are then given by
\begin{equation}
\label{eigen3}
s_{1,2}=u+\hat\beta(u^2+2\rho)\mp (1+2\hat\beta u)\sqrt{\rho}
\end{equation}
which give the characteristic velocities of the Riemann invariants $\la_1$
and $\la_2$ in (\ref{l0}), i.e. $\la_{1,2}=u\mp 2\sqrt{\rho}$.  In terms
of the Riemann invariants, Eqs.(\ref{eigen3}) are expressed as
\begin{equation}
\begin{array}{rcl}
& & s_1=\displaystyle{\frac{1}{4}(3\la_1+\la_2)+\frac{\hat \beta}{8}
(5\la_1^2+\la_2^2+2\la_1\la_2)}, \\
& & s_2=\displaystyle{\frac{1}{4}(\la_1+3\la_2)+\frac{\hat \beta}{8}
(\la_1^2+5\la_2^2+2\la_1\la_2)},
\end{array} \label{speed3}
\end{equation}
Because of the nonlinear terms of $\la_1$ and $\la_2$,
the characteristic velocity does not satisfy the double monotonicity in
Corollary \ref{coro1}.  This implies that the $g=0$ Whitham equation (\ref{g0t})
develops a shock even in the case where the Riemann invariants are both
decreasing functions.  To demonstrate the situation, we
 study the initial value problem of (\ref{g0t})
with the initial data given by (\ref{ig0}).
As in the previous section, we calculate the characteristic velocities of the
pulse edges at $T=-T_0$ and at $T=T_0$
(see Fig.\ref{f2:2}).

Let us first compute the velocity near $T=T_0$, where the Riemann invariant
$\la_2$ decreases from $2\sqrt{\rho_0}$ to $-2\sqrt{\rho_0}$, and $\la_1
=-2\sqrt{\rho_0}$.  From Eq.(\ref{speed3}), $s_2$ near $T=T_0$ is given by
\begin{equation}
\label{s2at0}
s_2(\la_1=-2\sqrt{\rho_0},\la_2)= \frac{5\hat\beta}{8}\la_2^2
+\frac{1}{4}(3-2\hat\beta\sqrt{\rho_0})\la_2 -\frac{\sqrt{\rho_0}}{2}
(1-\hat\beta\sqrt{\rho_0}).
\end{equation}
To avoid a shock, we require that $s_2$ is a monotone increasing function in
$\la_2\in[-2\sqrt{\rho_0}, 2\sqrt{\rho_0}]$.  This implies that for
$\hat\beta>0$
the velocity  $s_2$ has to take the minimum in the region
$\la_2\le-2\sqrt{\rho_0}$,
and we have the relation,
\begin{equation}
\frac{\partial s_2}{\partial \la_2}\Big|_{\la_2=-2\sqrt{\rho_0}}=
-3\hat\beta\sqrt{\rho_0}+\frac{3}{4}\ge 0.
\label{bp}
\end{equation}
Similarly, for $\hat\beta<0$, $s_2$ has to take the maximum
in $\la_2\ge2\sqrt{\rho_0}$, and
\begin{equation}
\frac{\partial s_2}{\partial \la_2}\Big|_{\la_2=2\sqrt{\rho_0}}=
2\hat\beta\sqrt{\rho_0}+\frac{3}{4}\ge 0.
\label{bm}
\end{equation}
 From these results we obtain
\begin{equation}
\label{bcond1}
-\frac{3}{8\sqrt{\rho_0}}\le \hat\beta \le \frac{1}{4\sqrt{\rho_0}}.
\end{equation}
To compute also the velocity $s_1$ near $T=-T_0$, we note the symmetry of
the Riemann invariant form of the $g=0$ Whitham equation (\ref{g0t}),
 that is, for the reflection $T\to -T$ the form is
invariant under the change of variables $\la_k \to -\la_{3-k}$ for $k=1,2$ and
$\hat\beta \to -\hat\beta$.  Therefore we obtain the condition to
avoid a shock near $T=-T_0$,
\begin{equation}
\label{bcond2}
-\frac{1}{4\sqrt{\rho_0}}\le \hat\beta \le \frac{3}{8\sqrt{\rho_0}}.
\end{equation}
Thus we have :
\begin{proposition}
The system (\ref{g03}) with the initial data (\ref{ig0})
admits a global solution if the following condition is satisfied,
\begin{equation}
\label{shock}
|\hat\beta| \equiv \Big|\frac{\beta_3}{(-\beta_2)^{3/2}}\Big|
\le \frac{1}{4\sqrt{\rho_0}}.
\end{equation}
\end{proposition}
This provides a condition on design parameters in the second and
third order dispersions and the power level for a smooth NRZ pulse propagation.

The solution of (\ref{g0t}) can be obtained as follows:
Following the proof of Proposition
\ref{prop1}, for example near $T=T_0$, we have a single equation for $\la_2$,
\begin{equation}
\frac{\partial \la_2}{\partial Z}=s_2(-2\sqrt{\rho_0},\la_2)
\frac{\partial \la_2}{\partial T},
\label{riemannt}
\end{equation}
where $s_2(-2\sqrt{\rho_0},\la_2)$ is given by (\ref{s2at0}).
The solution of this equation can be obtained in the hodograph form,
\begin{equation}
\label{hodo3}
T+s_2(-2\sqrt{\rho_0},\la_2) Z=T_0,
\end{equation}
from which $\la_2$ gives a monotone decreasing function in $T$ from $\la_2=
2\sqrt{\rho_0}$ to $\la_2=-2\sqrt{\rho_0}$, if the condition (\ref{shock})
is satisfied.  In Fig.\ref{f5:1}, we plot
the pulse profiles $\rho(T,Z)$ at $Z=5\beta\approx 1.58$ obtained by the
numerical
simulation of (\ref{hnls3}), where we used $\beta_2=-0.1$ and $\beta_3=0.02$,
giving a case of shock formation i.e. $\hat \beta=0.63>0.25$ with $\rho_0=1$.
In this example, we predict the shock formation from both edges
$T=\pm T_0$, which are due to the violation of the conditions (\ref{bcond1})
and (\ref{bcond2}).
\begin{figure}
\epsfysize=7cm
\centerline{\epsffile{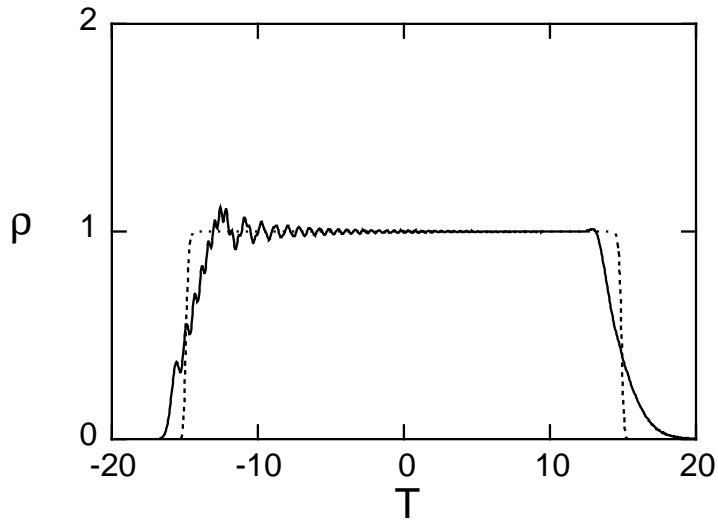}}
\caption{Pulse profile under the effect of third order dispersion.}
\end{figure}
The asymmetric deformation of the pulse
due to the third order dispersion has been observed in the experiment
\cite{akiba:91}

It is also interesting to study how one can regularize a case where the
condition (\ref{shock}) is violated.  For example, if
$\hat\beta=3/(2\sqrt{\rho_0})$,
then $s_2$ takes minimum at $\la_2=0$.  So $\la_2$ becomes multi-valued
function in $T$ for $Z>0$, which indicates a shock formation, and
one needs to regularize the initial data by choosing an appropriate genus.
This will be further discussed elsewhere.

\section{WDM-NRZ system}
\renewcommand{\theequation}{6.\arabic{equation}}\setcounter{equation}{0}

In a WDM system, signal in each channel has a different carrier frequency,
and the model equation of the system can be obtained by setting the electric
field $q$ of the NLS equation (\ref{nls}) in the form,
\begin{equation}
\label{wdm}
q=\sum_{j=1}^N q_j.
\end{equation}
Here $q_j$ represents the electric field in the $j$-th channel having
carrier frequency, say $\Omega_j$.  Because of the difference in
the carrier frequencies, pulse in one channel has
different group velocity from those of the others, and we see pulse-pulse
collision between channels.  As a simple but fundamental case,
we consider here the 2-channel
WDM system.  Then substitutng (\ref{wdm}) into the NLS equation (\ref{nls}), and
ignoring the mismatch of the frequency in the FWM terms, we obtain the following
coupled NLS equation \cite{mollenaure:90},
\begin{equation}
\left\{
\begin{array}{rcl}
 & {} & \displaystyle{i\frac{\partial q_1}{\partial Z}-{\beta^2 \over 2}
\frac{\partial^2 q_1}{\partial T^2} +  (|q_1|^2 + \alpha |q_2|^2) q_1=0}, \\
& {} & \\
 & {} & \displaystyle{i\frac{\partial q_2}{\partial Z}-{\beta^2 \over 2}
\frac{\partial^2 q_2}{\partial T^2} + (|q_2|^2 + \alpha |q_1|^2) q_2=0},
\end{array}
\right.
\label{cnls}
\end{equation}
where $\alpha=2$.  The validity of this model has been discussed in
\cite{kodama:96b}
where a dispersion management technology is used to reduce the
FWM and to keep a small GVD.  We should also
like to mention that the system (\ref{cnls})
with $\alpha=1$ is integrable, and is called the Manakov equation
\cite{manakov:78}.

Let us first summarize the result in \cite{kodama:96b} for the 2-channel WDM
case ($\alpha=2$).  The main result in \cite{kodama:96b} is that there
exists a critical frequency separation $\Delta \Omega_c$ such that
the system with the channel
separation $|\Omega_1-\Omega_2|<\Delta\Omega_c$ leads to a hydrodynamic-type
instability.  It is also interesting to note that this instability
appears even in the case of $\alpha=1$, and is
not essential for the specific value of $\alpha$.  So we study this instability
based on the Manakov equation, which enables us to characterize the
instability as a deformation of the branch points on
 the Riemann surface defined by the spectral
curve of the Lax operator for the equation.

The Lax operator $L_M$ for the Manakov equation takes a similar form
as of the NLS equation, (\ref{l}), and is given by
\begin{equation}
\label{lm}
L_M=D+\sqrt{\rho_1}(D-u_1)^{-1}\sqrt{\rho_1}+
\sqrt{\rho_2}(D-u_2)^{-1}\sqrt{\rho_2},
\end{equation}
where $\rho_k$ and $u_k$ are of course defined by
\begin{equation}
\label{qk}
q_k(T,Z)=\sqrt{\rho_k(T,Z)}\exp\left(i\frac{1}{\beta} \sigma_k(T,Z)\right),
\ \ \mbox{for} \ \ k=1,2.
\end{equation}
As seen from this formula, the Manakov equation can be easily extended to
an integrable $N$-coupled NLS equation which may correspond
to a model equation of $N$-channel WDM system.  Then the Manakov
equation can be written in the Lax form (\ref{lax}) with $B_2=(L_M^2)_+/2$, i.e.
\begin{equation}
\label{manakov}
-i\beta{\partial L_M \over \partial Z}=[B_2, L_M],
\end{equation}
where we have rescaled $Z$ to $Z/\beta$.

Now taking the limit $\beta \to 0$ and ignoring all the terms including
$\beta$, we obtain a hydrodynamic-type equation,
\begin{equation}
{\pat \over {\pat Z}} \left(
\matrix{
 \rho_1 \cr  u_1 \cr \rho_2 \cr u_2 \cr}
 \right)  = \left(
 \matrix{
 u_1 & \rho_1 & 0 & 0 \cr
 1 &  u_1 & \alpha & 0\cr
  0 & 0 & u_2 & \rho_2 \cr
  \alpha & 0 & 1 & u_2 \cr}
 \right)
{\pat \over {\pat T}} \left(
\matrix{
\rho_1 \cr  u_1 \cr \rho_2 \cr u_2 \cr} \right).  \label{g03}
\end{equation}
Note that this equation can also be derived from the original system
(\ref{cnls})
with arbitrary $\alpha$ by substituting (\ref{qk}) into (\ref{cnls}) in the
same approximation.  However a system with $\alpha\ne 1$ may not be integrable
and the corresponding Whitham equation may not be constructed in a
systematic manner.

To see the hyperbolicity of (\ref{g03}), we first calculate the characteristic
polynomial of the coefficient matrix in (\ref{g03}),
\begin{equation}
\label{chara}
[(u_1-s)^2-\rho_1][(u_2-s)^2-\rho_2] -\alpha^2 \rho_1\rho_2=0.
\end{equation}
As a practical example,
 we set the equal intensity $\rho_1=\rho_2\equiv \rho_0$ and the opposite
frequency shifts $u_1=-u_2\equiv u_0$, then we have a condition for the
hyperbolicity of (\ref{g03}), that is, all the eigenvalues of the
coefficient matrix are real, if the following condition is satisfied,
\begin{equation}
\label{du}
|u_0| > [(1+\alpha)\rho_0]^{1/2}.
\end{equation}
This then gives a minimum frequency separation for 2-channel WDM system
for a stable NRZ pulse propagation.  The instability corresponding to the
complex eigenvalues (characteristic velocity)
is called hydrodynamic-type instability in \cite{kodama:96b}.
Thus the instability is due to the change of type of the system (\ref{g03})
from hyperbolic to elliptic.

We now consider the geometry
 of the instability using the Manakov equation, and discuss the associated
 Whitham equation to describe a schock formation in WDM system.
In the limit $\beta \to 0$, we again assume the wave function $\phi$ of
$L_M\phi=\lambda\phi$ in the WKB form (\ref{wkb}), then the spectral
curve $\la=\la(P)$ is given by
\begin{equation}
\label{lp}
\la=P+{\rho_1 \over P-u_1} + {\rho_2 \over P-u_2},
\end{equation}
where $P=\partial S/\partial T$, and the evolution of $P$ is given by
$\partial P/\partial Z=\partial Q_2/\partial T$ with $Q_2=(1/2)[\la^2]_{\ge 0}=
P^2/2 +\rho_1+\rho_2$, the limit of $B_2\phi/\phi$ in (\ref{manakov}).
\begin{figure}
\epsfysize=7cm
\centerline{\epsffile{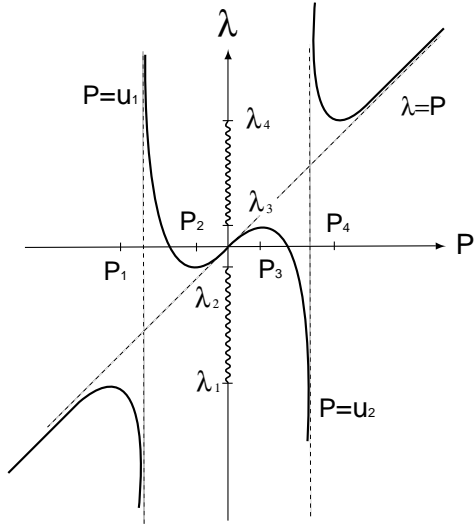}}
\caption{The $g=0$ algebraic curve (6.9).}
\end{figure}
In Fig.\ref{f6:1}, we plot the graph of Eq.(\ref{lp}) on the $P$ - $\la$ plane.
As we see from Fig.\ref{f6:1}, there are at most three
 roots of $P$ in Eq.(\ref{lp}) for each $\la$, defining the three sheeted
 (non-hyperelliptic) Riemann surface.
The characteristic polynomial (\ref{chara}) is then given by the equation
$\partial \la/\partial P=0$, and the eigenvalues are obtained from (\ref{lp})
at the roots of this equation.  Thus the spectral of $L_M$ of (\ref{lm}) is
given by
\begin{equation}
\label{specm}
\mbox{Spec}(L_M)=(-\infty, \la_1]\cup[\la_2, \la_3]\cup[\la_4, \infty).
\end{equation}
We also note that the spectral curve (\ref{lp}) defines the $g=0$ Riemann
surface whose topology is a sphere obtained
by a compactified $P$-plane (see Fig.\ref{f6:2}).

\begin{figure}
\epsfysize=5.5cm
\centerline{\epsffile{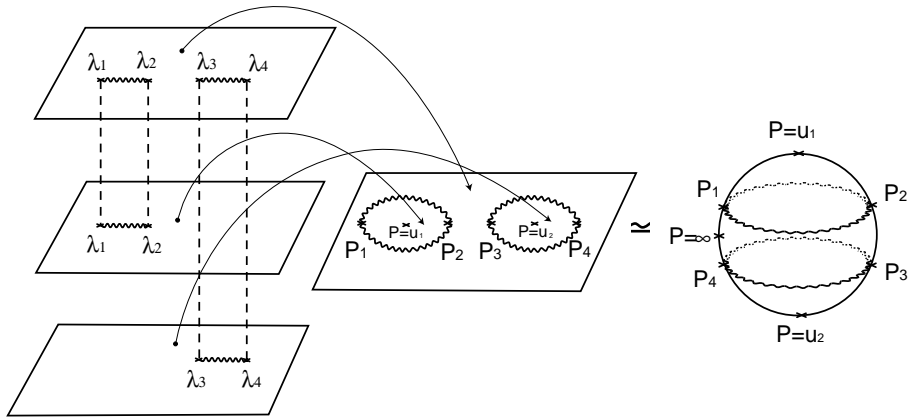}}
\caption{The $g=0$ Riemann surface corresponding to the curve (6.9).}
\end{figure}

To see these more precisely, let us study a detail of the algebraic curve
(\ref{lp}).  We first note
 Eq.(\ref{lp}) can be written as
\begin{equation}
\label{nhcurve}
4y^3 -3 \alpha_0 (\la) y - \beta_0 (\la) = 0,
\end{equation}
where $2y:=3P-(\la+u_1+u_2)$, and $\alpha_0, \beta_0$ are
the monic polynomials of degree 2 and 3
in $\la$,
\begin{eqnarray}
& & \alpha_0=\displaystyle{{\la^2 }-(u_1+u_2)\la+u_1^2+u_2^2-u_1u_2-3(\rho_1+
\rho_2),}  \label{a0} \\
& & \beta_0=\displaystyle{\la^3-{3\over 2}(u_1+u_2)\la^2-{3\over
2}\left\{u_1^2+u_2^2-4u_1u_2
+3(\rho_1+\rho_2)\right\}\la} \nonumber \\
& & ~~~~~~\displaystyle{-{9 \over
2}(\rho_1u_1+\rho_2u_2)+9(\rho_1u_2+\rho_2u_1).}
 \label{b0}
\end{eqnarray}
The roots of (\ref{nhcurve}) are given by
\begin{equation}
y=\frac{1}{2}\left(\omega^{\ell}\sqrt[3]{\nu_+}
+\omega^{3-\ell}\sqrt[3]{\nu_-}\right),
~~~~~~~ \mbox{for} ~~~~\ell=0,1,2, \label{root}
\end{equation}
where $\omega=\exp(2\pi i/3)$, and $\nu_{\pm}$ are given by
\begin{equation}
\nu_{\pm}=\beta_0\pm\sqrt{\beta_0^2-\alpha_0^3}.
\end{equation}
Each root lives in a different sheet, and in particular
the root marked by $\ell =0$ in (\ref{root}) lives in the fundamental sheet
having the property $P\to \la$ as $\la\to \infty$. With this root,
Eq.(\ref{g03})
can be identified as the $g=0$ Whitham equation in the form (\ref{whitham})
where the differentials $\omega_1$ and $\omega_2$ are given by
\begin{eqnarray}
& &\omega_1=dP=\frac{1}{3}\left(1+\frac{1}{3(\nu_+)^{2/3}}{d\nu_+ \over d\la}
+\frac{1}{3(\nu_-)^{2/3}}{d\nu_- \over d\la}\right) d\la \nonumber \\
& & ~~~~~~~~~~~~~ \sim
\left(1+\frac{\rho_1+\rho_2}{\la^2}+\cdots\right)d\la, \label{o1} \\
& & \omega_2=dQ_2=PdP \nonumber \\
& & ~~~~~~~~~~~~~ \sim \left(\la
+\frac{u_1\rho_1+u_2\rho_2}{\la^2}+\cdots\right)d\la. \label{o2}
\end{eqnarray}
 The branch points (Riemann
invariants) $\la_k, k=1,\cdots,4$ are obtained from the descriminant
$\Delta_0(\la)$
of the equation (\ref{nhcurve}),
that is,
\begin{eqnarray}
 & & \Delta_0\equiv \displaystyle{-\frac{4}{27}(\beta_0^2-\alpha_0^3)} \nonumber \\
 & & ~~~~~~=\displaystyle{(u_1-u_2)^2\prod_{k=1}^4(\la-\la_k)=0}. \label{desc}
\end{eqnarray}
 In the case where
$u_1=-u_2\equiv u_0$ and $\rho_1=\rho_2\equiv \rho_0$, the roots of
Eq.(\ref{desc}) are given by
\begin{equation}
\label{lamda}
\la^2=\frac{1}{u_0^2}\left\{2u_0^4+10u_0^2\rho_0-\rho_0^2
\pm(4u_0^2+\rho_0)^{3/2}\sqrt{\rho_0}\right\}.
\end{equation}
We also culculate the characteristic
velocities $s_k=P_k\equiv P(\la_k)$ which is given by
the roots of $\partial \la/\partial P=0$,
\begin{equation}
P^2=u_0^2+\rho_0\pm\sqrt{(4u_0^2+\rho_0)\rho_0}.
\label{pcha}
\end{equation}
If $u_0^2>2\rho_0$ of (\ref{du}), these roots are all real.
Note that $\la_k=\la(P_k)$ with the orderings $\la_1<\la_2<\la_3<\la_4$ and
$P_1<P_2<P_3<P_4$.  Then the condition (\ref{du}) indicates the non-closing
of the interval [$\la_2,\la_3$].

Now let us consider the case with nonzero genus.  The corresponding
algebraic curve
may be given  by
\begin{equation}
\label{curve3}
4y^3 -3\alpha_h(\la) y - \beta_h(\la) =0,
\end{equation}
with
\begin{eqnarray}
\begin{array}{ll}
& \alpha_h=\displaystyle{\prod_{j=1}^{2h+2}(\la-\xi_j),} \\
& \beta_h = \displaystyle{\prod_{k=1}^{3h+3}(\la-\eta_k),}
\end{array}
\label{hcurve}
\end{eqnarray}
where we have the relation $3\sum_{j=1}^{2h+2}\xi_j=2\sum_{k=1}^{3h+3}\eta_k$.
Then the descrimenant $\Delta_h:=-(4/27)(\beta_h^2-\alpha_h^3)$ of the curve
(\ref{curve3}) gives a
polynomial of degree $6h+4$ in $\la$, and it indicates that there
are at most $3h$ numbers of
genus openning in the Whitham equation.
Note that the genus of the (non-hyperelliptic)
Riemann surface defined by the curve (\ref{curve3}) is generically given by
\begin{equation}
\label{gnumber}
g=3h.
\end{equation}
The number 3 implies that gap-opening appears in between 2 sheets of
 the total 3 sheets.  A further discussion on this subject will be given
 elsewhere.

 \vspace{1cm}

\noindent
{\bf Acknowledgement.}
The author would like to thank Ann Morlet for a useful discussion
on the begining
stage of this paper, and Akihiro Maruta for making several figures
obtained by the numerical simulations of the NLS equation.  The work is
partially
supported by NSF grant DMS9403597.

\newpage
\Large
{\bf Figure Captions\\}
\normalsize
\begin{enumerate}
\baselineskip 25 pt

\item \label{f2:1}
Various types of a 16 bits-coded signal (0010110010111100).

\item \label{f2:2}
Evolution of the Riemann invariants.  Doted curve are the initial data;
Solid curves are the solutions $\la_k(T,Z)$ for $Z>0$.

\item \label{f2:3}
Deformation of NRZ pulse (Dam breaking problem).  The dotted curve in $\rho$
shows the numerical result of the NLS equation.

\item \label{f2:4}
Optical shock due to initial chirp.

\item \label{f3:0}
The $g=0$ algebraic curve (\ref{genus0}).

\item \label{f3:01}
The $g=0$ Riemann surface corresponding to the curve (\ref{genus0}).

\item \label{f3:02}
The $g=1$ algebraic curve $y^2=R_g(\la)$ in (\ref{genusg}).

\item \label{f3:1}
The 2-sheeted Riemann surface with branch cuts and canonical cycles.

\item \label{f4:1}
The function $V_i(\la)$ with $i=2k +1$.

\item \label{f4:1e}
The initial phase modulation for one-bit pulse.

\item \label{f4:2}
Evolution of NRZ pulse with initial phase modulation.  a) Without, and b)
with the initial modulation.

\item \label{f4:i1}
Evolution of the Riemann invariants for $u_0>2\sqrt{\rho_0}$.

\item \label{f4:i2}
Deformation of the pulse $\rho(T,Z)$ for $ u_0>2\sqrt{\rho_0}$.

\item \label{f4:ii}
Evolution of the Riemann invariants for $ 0<u_0<2\sqrt{\rho_0}$.

\item \label{f4:iii1}
Evolution of the Riemann invariants for $ 0>u_0>-2\sqrt{\rho_0}$.

\item \label{f4:iii2}
Deformation of the pulse $\rho(T,Z)$ for $0> u_0>-2\sqrt{\rho_0}$.
Solid curve, the numerical result; Dashed curve, the solution (4.34).

\item \label{f4:iv1}
Evolution of the Riemann invariants for $ u_0<-2\sqrt{\rho_0}$.

\item \label{f4:iv2}
Deformation of the pulse $\rho(T,Z)$ for $ u_0<-2\sqrt{\rho_0}$.

\item \label{f4:5}
Deformation of the Riemann surface for  $ 0<u_0<2\sqrt{\rho_0}$.

\item \label{f5:1}
Pulse profile under the effect of third order dispersion.

\item \label{f6:1}
The $g=0$ algebraic curve (\ref{lp}).

\item \label{f6:2}
The $g=0$ Riemann suface corresponding to the curve (\ref{lp}).

\end{enumerate}

\end{document}